\documentclass[final,3p,times,twocolumn]{elsarticle}

\usepackage{lineno,hyperref}

\usepackage[english]{babel}
\usepackage{amsmath}  
\usepackage{amsfonts} 
\usepackage{graphicx} 
\usepackage{xcolor}
\usepackage{siunitx}
\usepackage{wasysym}
\usepackage[utf8]{inputenc}
\usepackage{lipsum}
\usepackage{booktabs}
\usepackage{stfloats}
\usepackage{caption}
\usepackage{capt-of}
\usepackage{stfloats}
\usepackage{makecell}
\usepackage{enumitem}
\usepackage{calc}
\usepackage{tabto}
\usepackage{subscript}
\TabPositions{5em}
\usepackage{doi}
\usepackage{bm}
\usepackage{mhchem}

\usepackage{caption}
\usepackage{subcaption}
\usepackage{flushend}
\usepackage{balance}

\begin{document}
%\pagewiselinenumbers
%\linenumbers
\sloppy
\begin{frontmatter}

\journal{journal}

\title{Neutron spectrometry of (p,n) based neutron fields at the University of Helsinki using the NCT-WES single-moderator directional spectrometer}

\author[HY,HIP]{M.\ Myllymäki\corref{corr}}
\ead{mila.myllymaki@helsinki.fi}
\author[HY,HIP]{S.\ Saariokari}
\author[HY]{A.\ Perkkiö}
\author[HIP]{H.\ Badran}
\author[INFN]{R.\ Bedogni}
\author[HY,HIP]{E.\ Brücken}
\author[INFN]{A.\ I.\ Castro Campoy}
\author[INFN]{L.\ Russo}
\author[INFN,ICTP]{F.\ L.\ Frigi}
\author[CIEMAT,INFN]{J.\ M.\ Gomez-Ros}
\author[HY]{K.\ Helariutta }
\author[HIP]{T.\ Hildén}
\author[STUK]{J.\ Huikari}
\author[INFN,ITA]{J.\ Lapa}
\author[INFN,ENEA]{A.\ Pietropaolo\corref{corr}} 
\ead{antonino.pietropaolo@enea.it}
\author[HIP,LUT]{M.\ Väänänen}
\author[HY,HIP,STUK]{T.\ Siiskonen}
\cortext[corr]{Corresponding authors}

\address[HY]{University of Helsinki, Department of Physics, P.O. Box 64, FI-00014 University of Helsinki, Finland}
\address[HIP]{Helsinki Institute of Physics, P.O. Box 64, FI-00014 University of Helsinki, Finland}
\address[LUT]{LUT University, P.O. Box 836, FI-00074, Finland}
\address[STUK]{STUK – Radiation and Nuclear Safety Authority, Vantaa, Finland}
\address[INFN]{Istituto Nazionale di Fisica Nucleare - Frascati National Laboratories, Frascati, Italy}
\address[ICTP]{ICTP International Centre for Theoretical Physics, Strada Costiera, 11, 34151 Trieste, Italy}
\address[PM]{Politecnico di Milano, Dipartimento di Energia, via La Masa 34, 20156 Milano, Italy}
\address[ENEA]{ENEA - Department of Fusion and Technologies for Nuclear Safety and Security, Frascati, Italy}
\address[CIEMAT]{CIEMAT, Av. Complutense 40, 28040 Madrid, Spain}
\address[ITA]{ITA - Technological Institute of Aeronautics, São José dos Campos, SP, Brazil}

\begin{abstract}
A tunable-energy neutron beam produced with up to 1 $\mu$A / 10 MeV proton beam on a solid LiF target is available at the Department of Physics of the University of Helsinki for neutron metrology and research applications. This work presents the experimental characterisation of the neutron beams produced at the Helsinki Accelerator Laboratory, and the validation of the \texttt{MCNP6} and \texttt{GEANT4} radiation transport codes for simulating these neutron beams. The neutron spectra produced in the forward direction from 1960, 2040, and 2310 keV protons were measured using the NCT-WES single-moderator directional spectrometer. NCT-WES (Neutron Capture Therapy Wide Energy Spectrometer) is a recently developed single-moderator neutron spectrometer with a sharply directional response, and it is a very convenient instrument to characterise neutron facilities, as it determines the whole spectrum in a single exposure, and has a low sensitivity to room-scattering. The neutron fluence was also determined with the Berthold LB 6411 neutron monitor. The data acquired with NCT-WES were unfolded using the FRUIT unfolding code, starting from guess spectra obtained from \texttt{GEANT4} or \texttt{MCNP6} simulations. The unfolding analysis showed that both transport codes produced plausible spectra, \texttt{MCNP6} being more accurate for the 1960 and 2040 keV irradiations and \texttt{GEANT4} for 2310 keV. The measured values of energy-integrated neutron fluence agree satisfactorily with simulations as well as preliminary measurements with the LB 6411 neutron monitor.
\end{abstract}

\begin{keyword}
Neutrons \sep lithium target \sep accelerator based neutron source \sep spectroscopy \sep Monte Carlo \sep GEANT4 \sep MCNP6.
\end{keyword}

\end{frontmatter}

\section{Introduction}

%\begin{figure}[h]
%\centering
%\includegraphics[width=0.3\textwidth,clip]{pics/bettan.png}
%\caption{\red{some image of the setup, tbd}}
%\label{fig:setup}
%\end{figure}

Well characterised neutron beams with narrow energy distribution are used for the calibration of neutron sensitive detectors or for irradiation of instruments or other targets. In many applications neutrons within a specific energy range are desirable. Neutrons can be produced via a variety of nuclear reactions, using either radioactive sources or accelerator-based approaches. Unlike radioactive sources such as \ce{^252Cf} or \ce{^241AmBe} that emit a broad neutron spectrum up to several MeV, accelerator-based fields can be set up to produce relatively narrow energy distributions \cite{nolte2011monoenergetic}. Using e.g.\,(p,n) reactions in neutron production, the maximum neutron energy can be accurately controlled by adjusting the incident proton energy relative to the reaction threshold. According to the International Organisation for Standardisation (ISO), monoenergetic neutron reference fields in the 100 to 600 keV range can be produced by the \ce{^7Li}(p,n) reaction in a lithium target using proton energies of 1940, 2020, and 2297 keV (reaction threshold 1880 keV) \cite{nolte2011monoenergetic}. The highest reference energy is above the broad resonance peak of the reaction at 2250 keV. These reference neutron fields can then be moderated to adjust the energy distribution to be suitable for a particular user application. 

Accelerator-based neutron sources are particularly promising for applications in medical physics, such as Boron Neutron Capture Therapy (BNCT), where epithermal neutron fields are required. The development of accelerator-based neutron sources has made it possible to produce neutron fields suitable for BNCT using accelerators installed in hospitals. Worldwide, several facilities are either operational, undergoing clinical trials, or currently being developed \cite{Kumada2023}. Protons are generally used as primary particles in these facilities, with lithium or beryllium as the target material. The resulting neutrons are then moderated to the epithermal energy range with a suitable beam shaping assembly (BSA). 

The development of accelerator-based neutron sources for BNCT also creates the need for appropriate detectors for beam characterisation, monitoring, and dosimetry. However, access to clinical BNCT facilities for detector development and testing is limited. To provide an accessible alternative for detector testing, a tunable accelerator-based neutron beamline at the Helsinki Accelerator Laboratory of the University of Helsinki was designed and characterised. This work presents the experimental and computational characterisation of three unmoderated neutron spectra generated from a lithium fluoride target using three different proton energies based on the ISO recommendations.

A thorough spectrometric characterisation of these neutron fields is crucial for further beam shaping as well as opening the facility for other user access programmes. As these neutron fields are expected to span over various energy decades, a wide energy interval neutron spectrometer is required for the characterisation. Although conventional Bonner Spheres (BS) \cite{pietropaolo2020neutron} have the desired energy response, their operation is very time consuming, as each sphere requires individual exposure at the measurement point. In addition, due to their isotropic response and the shape of their response matrix, BS are highly sensitive to the low-energy room scatter component, eventually requiring very long measurements with shadow-cones \cite{iso8529_2} to correct for this effect. Single moderator neutron spectrometers (SMNS) \cite{bedogni2015experimental,bedogni2019thermal,bedogni2019cysp} are very similar to BS in terms of energy response. However, since they include an array of parallelised thermal neutron sensors, they need a single exposure to determine the whole neutron spectrum. Moreover, SMNS, based on a collimated moderating cylinder \cite{bedogni2015experimental,bedogni2019thermal,bedogni2019cysp,bedogni2020design,gomez2015cysp}, exhibits a sharply directional response and can be conveniently used to suppress room-scattered neutrons in scenarios where they constitute a parasitic effect, as in this work. Among the different cylindrical SMNS developed by the Frascati group, NCT-WES (Neutron Capture Therapy Wide Energy Spectrometer) \cite{bedogni2021nct,bedogni2023nct} is the most indicated for the fields studied in this work, as it was especially designed to offer maximum resolving power in the energy region 0.05 – 1 MeV, where these fields show most of their fluence. This device was already successfully used for the spectrometric characterisation of metrology-grade neutron irradiation facilities \cite{bedogni2024characterisation,bedogni2025neutron}.

 Monte Carlo simulations are a vital tool for designing neutron beamlines as they allow evaluating neutron beam characteristics such as the energy spectrum and the flux without time-consuming measurements. Consequently, validating these simulation tools against experimental results is necessary for further beamline optimisation and design. In this work, the design and characterisation of the beamline is presented. Additionally two common Monte Carlo codes, \texttt{MCNP6} and \texttt{GEANT4}, are compared and validated against the NCT-WES measurements, using a second detector to obtain an independent cross-check of the neutron flux.

\section{Experimental set-up}

\subsection{Detectors}
\begin{figure}[h]
\centering
\includegraphics[width=0.48\textwidth]{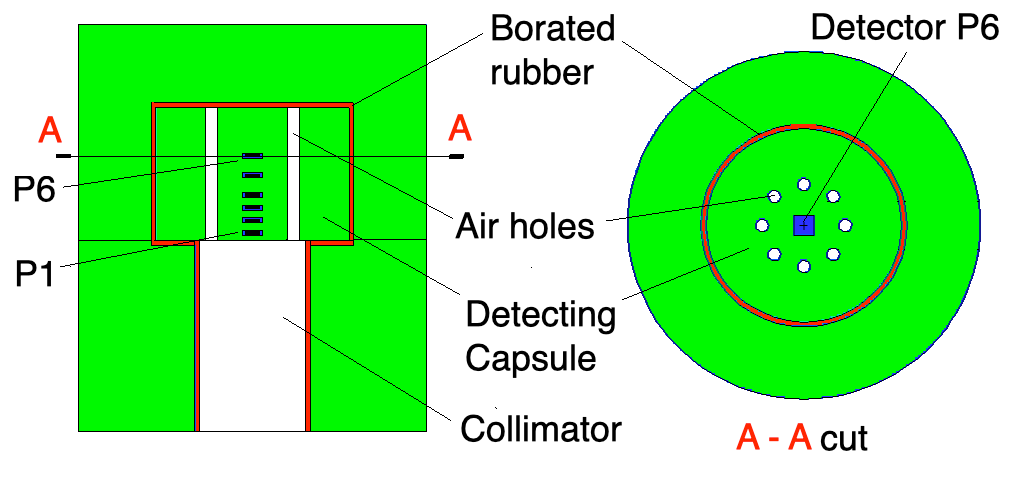}
\caption{Scheme of NCT-WES. Labels P1 and P6 indicate the shallowest and the deepest detector, respectively.}
\label{fig:nct-wes}
\end{figure}

Although full details on NCT-WES construction are given elsewhere \cite{bedogni2021nct}, a few design elements and a simplified scheme (Figure \ref{fig:nct-wes}) are given here to facilitate the reader. The diameter of NCT-WES is 36 cm and its length is 41.5 cm. Six TNPD-type \cite{bedogni2015experimental_semicon,bedogni2021modelling} thermal neutron sensors are positioned in the detection capsule along the cylindrical axis. The frontal collimator (20 cm long, 11 cm diameter, with 0.5 cm internal borated rubber lining) is pointed toward the direction under study, allowing the sensitive capsule to be irradiated along the desired direction. Neutrons from other directions, such as those composing the room scattering, tend to be moderated and absorbed in the lateral shield formed by polyethylene and borated rubber. The distance from the front face of the spectrometer and the centres of the detectors is 20.22\,cm (shallowest detector in position P1), 21.52\,cm (position P2), 22.82\,cm (position P3), 24.12\,cm (position P4), 26.12\,cm (position P5), and 28.12\,cm (deepest detector in position P6). The detection capsule includes eight air cavities to enhance the epithermal response. The electronic readout and data acquisition process is given elsewhere \cite{bedogni2021nct}.

\begin{figure}[h]
\centering
\includegraphics[width=0.48\textwidth]{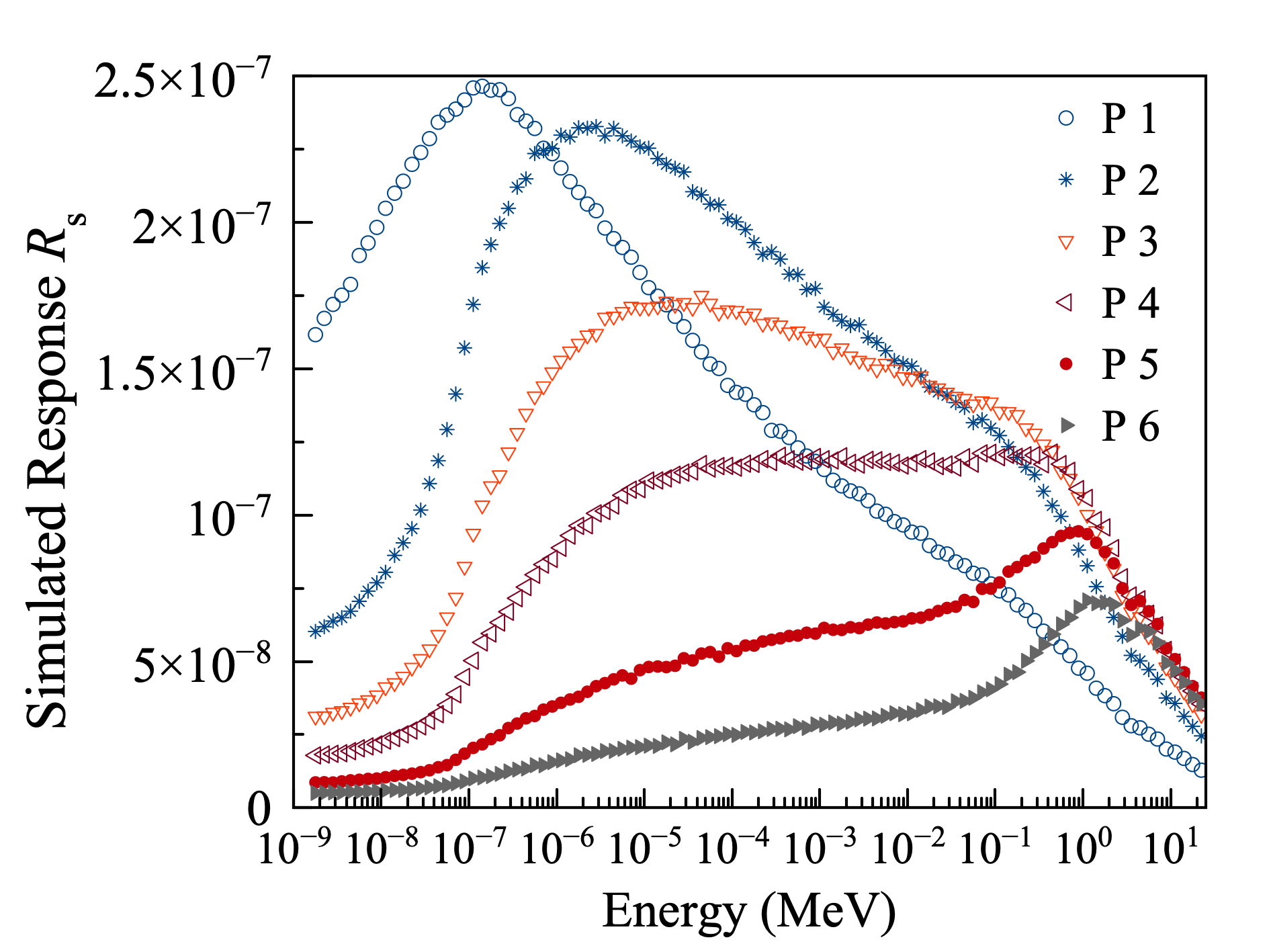}
\caption{Simulated Response of NCT-WES distance 138 cm, in terms of expected number of pulses per emitted neutron. Symbols P1 to P6 indicate the detector positions from the shallowest to the deepest.}
\label{fig:sim_res_nct-wes}
\end{figure}

The NCT-WES computational model was elaborated with the Monte Carlo N-Particle transport code (\texttt{MCNP6}) \cite{kulesza2022mcnp}, using the \texttt{ENDF/B-VIII} \cite{brown2018endf} neutron cross-section libraries below 20 MeV and room-temperature cross-section tables in polyethylene, S($\alpha$,\,$\beta$). This model was previously validated with monoenergetic reference fields at the National Physical Laboratory (NPL), and its overall uncertainty was estimated to be lower than ±2\,\% \cite{bedogni2023nct}.
The simulated response of the spectrometer is defined as the number of expected signals in the internal detectors per unit source neutron, as a function of the detector position and neutron energy, when its front-face is exposed to the neutron source. When the distance from the source is very large compared to the size of the spectrometer, its front face is uniformly irradiated. Thus, a response matrix derived under a broad parallel beam can be used. In contrast, when the distance from the source is comparable to the size of the spectrometer, a distance-dependent response matrix must be derived. This is the case in this work, where the distance from the centre of the target and the front face of the spectrometer was 138 cm. The associated response matrix is shown in Figure \ref{fig:sim_res_nct-wes}. Shallow detectors (P1 to P3) are mainly sensitive to low energies, as their responses tend to peak below 10 eV. Deeper detectors exhibit maxima in the 0.1 – few MeV region. As in the case of BS, the energy resolution is limited by the partial overlap and similarities in the shape of the response functions of the different detectors \cite{reginatto2002resolving}. The resolving power tends to be lower in energy regions where no maxima are featured, as from 0.1 to 10\,keV, and increases from 10\,keV to a few MeV. 

The Berthold LB 6411 neutron monitor \cite{Berthold} was used to determine the neutron fluence for each neutron field. The LB 6411 monitor consists of a high-density polyethylene moderator sphere, with a diameter of 25 cm. A composite \ce{^3He} recoil proton counter tube is located in the centre of the moderator sphere. The measurable neutron energy range of the LB 6411 monitor extends from thermal energies up to 20 MeV, and it has a high sensitivity of approximately 3 pulses per nSv, making it suitable for measuring low-intensity neutron fields. The LB 6411 is read out through a dose rate monitor that is permanently attached to the device. The readings are transferred to the control room with a CCD camera.

\subsection{Irradiation facility and measurement set-up}
The proton beams were produced with the EGP-10-II accelerator of the Helsinki Accelerator Laboratory at the University of Helsinki. EGP-10-II is a tandem van de Graaff ion accelerator with a maximum terminal voltage of 5 MV.  

\begin{figure}[h]
\centering
\includegraphics[width=0.47\textwidth]{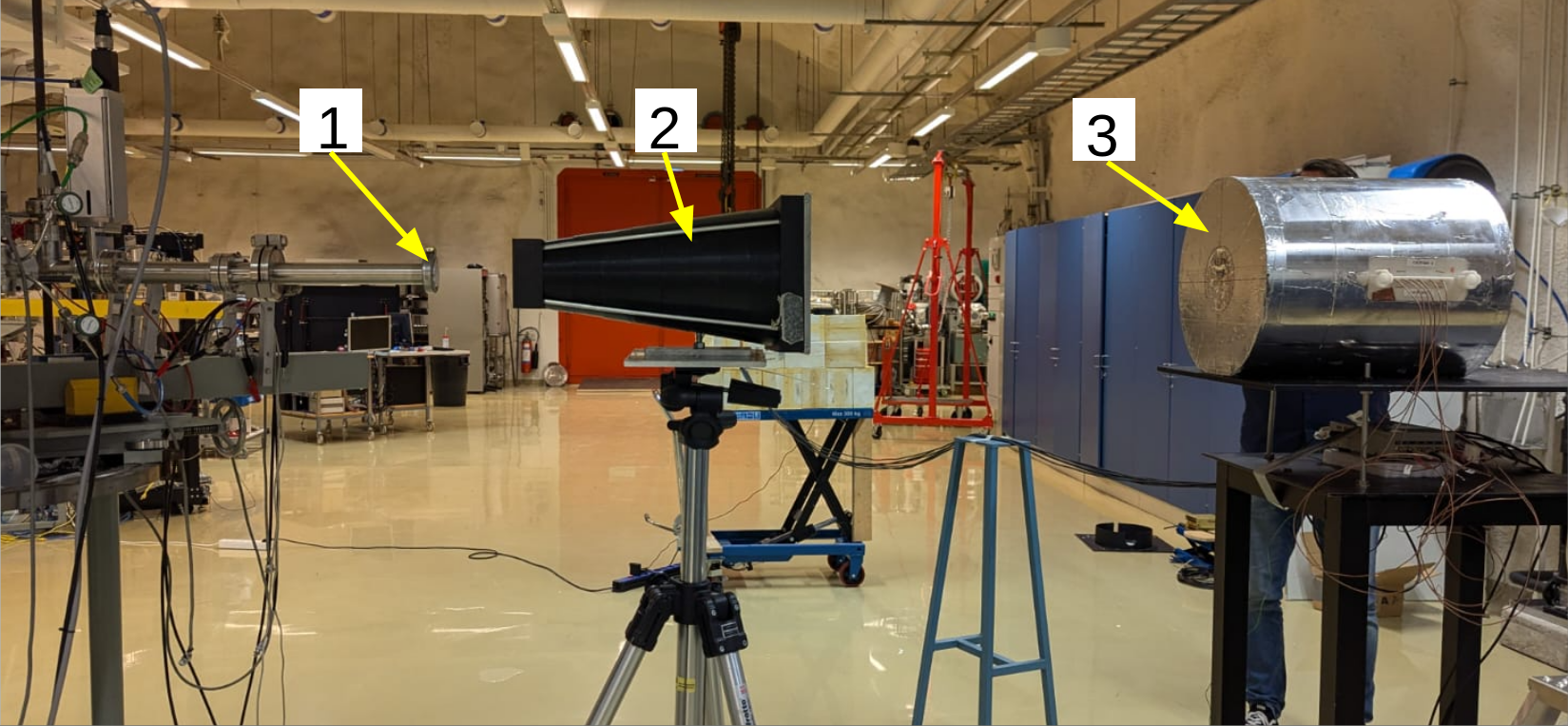}
\caption{Experimental set-up: (from left) neutron producing target (1), shadow-cone (2), NCT-WES spectrometer (3).}
\label{fig:experimental_set-up}
\end{figure}

The LiF target was installed at the end of the central beam line in the experimental hall of the accelerator. The experimental setup was built behind the target (Figure \ref{fig:experimental_set-up}). The experimental hall is a large room with concrete and rock walls, floor, and ceiling. The nearest walls to the spectrometer were the floor (1.8 m from the spectrometer), the back wall (2.5 m) and the ceiling (3.3 m). The distance to the other walls was at least 15 m. Neutron beams were produced with 1960, 2040, and 2310 keV protons. These values correspond to the ISO recommendations for producing mono-energetic neutron fields from lithium targets, however, a small offset was included in each proton energy to compensate for the accelerator uncertainties. The standard uncertainty in proton energy was estimated to be 5 keV. Proton beam current was monitored with a Faraday cup.
%The roof was quite low (3.3 m from the spectrometer), being the main source of room scattering together with the floor (1.8 m from the spectrometer). The distance between the spectrometer and the walls was 2.5 m (back-wall) and at least 15 m (other walls). Neutron beams produced with 1960, 2040, and 2310 keV protons were studied. 

The neutron production target was a natural LiF cylinder sold commercially for optical applications. It had a thickness of 4~mm and a diameter of 38.1~mm, and it was attached to the stainless steel beam pipe using an aluminium flange with a small insert to hold the target tightly in place. A small amount of aluminium foil was inserted between the sides of the LiF target and the flange to secure it. It should be noted that the target used is considered to be a thick LiF target, in which proton energy loss causes an energy spread in the produced neutron fields. Therefore, the resulting neutron spectra are not expected to be mono-energetic, unlike the ISO reference neutron fields.

%The Southern Scientific KWD 2222A Digipig neutron rem-meter was placed at about 4 m from the NCT-WES at about a 45° angle with respect to the proton beam to monitor the time-dependent neutron emission from the target. 

The neutron fluence for each neutron field was also determined with the Berthold LB 6411 neutron monitor. The centre of the LB 6411 was set at the reference point of the NCT-WES spectrometer (138 cm from the centre of the target). For each proton energy, the LB 6411 monitor was irradiated for 300 s, and the reading was read from the monitor screen every 60 s. The known energy response of the LB 6411 monitor \cite{Burgkhardt1997} was used to estimate the neutron fluence at the measurement point, based on the measured neutron energy spectra. 

The shadow cone technique \cite{iso8529_2} was used to correct the NCT-WES and LB 6411 readings for room scattering. The shadow cone is made of high-density polyethylene and has an aluminium support structure. The cone length is 50 cm and the end dimensions are 12x12 cm$^2$ and 26x26 cm$^2$. The larger end of the cone is covered with 1 mm of cadmium. When used, the smaller end of the cone was placed 15 cm from the target. 

Table \ref{tab:exp_data_summary_nctwes} presents a summary of the experimental data for the NCT-WES measurements: the measurement times, specified for free field or shadow-cone measurements, and the variability range for the scattered fraction ($s$ = shadow measurement / free field counting) over the NCT-WES internal detectors. As expected, $s$ tends to increase with energy. Nevertheless, due to the sharply directional response of the spectrometer, its value is always lower than about 3\%. The data of irradiation for the LB 6411 measurements are given in Table \ref{tab:exp_data_summary_berthold}.

\begin{table*}[t]
    \centering
    \caption{Data of the irradiation for the NCT-WES measurements: mean proton current, measurement times, and variability range for the scattered fraction $s$ over the NCT-WES internal detectors.}\label{tab:exp_data_summary_nctwes}
    \begin{tabular}{lccc}\toprule
        Proton energy [keV] & 1960 & 2040 & 2310 \\\midrule
        Proton current, free field [$\mu$A]& 0.430 $\pm$ 0.001 & 0.146 $\pm$ 0.002 & 0.135 $\pm$ 0.002\\
        Proton current, shadow cone [$\mu$A]& 0.343 $\pm$ 0.001 & 0.259 $\pm$ 0.001 & 0.576 $\pm$ 0.002\\
        Measurement time, free field (h) & 1.3 & 1.8 & 2.0 \\
        Measurement time, shadow cone (h) & 1.9 & 2.6 & 3.6 \\
        Range of scattered fraction $s$ & 1.2\% - 2.3\% & 0.4\% - 1.8\% & 1.9\% - 3.3\% \\\bottomrule
    \end{tabular}
\end{table*}

\section{Simulating the neutron production in the LiF target}
\label{sec:simulations}

Neutron production from a thick natural LiF target was simulated using \texttt{MCNP6} (version 6.3.0)\,\cite{kulesza2022mcnp} and \texttt{GEANT4} (version 11.2.2)\,\cite{Agostinelli:2002hh,Allison:2016lfl} transport codes. 

The simulation geometry included the vacuum-filled stainless steel beam pipe with an inner diameter of 5 mm, the 4 mm thick LiF target (diameter 38.1 mm, $\rho=2.635$ g/cm$^3$), and the 2 mm thick aluminium end flange ($\rho=2.7$ g/cm$^3$). The stainless steel beam pipe was modelled as SS316L ($\rho=8.00$ g/cm$^3$). The target was modelled with Li having an isotopic composition of 92.41\% \ce{^7Li} and 7.59\% \ce{^6Li}. The neutron production from the target is dominated by the \ce{^7Li(p,n)} reaction, since the threshold for \ce{^6Li(p,n)} reaction (5.9 MeV) is above the proton energies used in this work. The simulation world was filled with dry air at NTP. The floor, ceiling, or walls were not included in the simulation geometry. Therefore, the simulated neutron spectra correspond to no-scatter conditions, i.e., zero scattered fraction below. 

The \texttt{ENDF70PROT} cross section library \cite{Trellue_2008}, based on \texttt{ENDF/B-VII.0}, was used for incident proton interactions in the \texttt{MCNP6} simulations. For neutron interactions, the \texttt{Lib80x} cross section library \cite{Conlin_2018}, based on \texttt{ENDF/B-VIII.0}, was used. In both cases, cross section data evaluated at room temperature (293.6 K) were utilized. The \texttt{ENDF70PROT} library does not contain cross section data for proton interactions in $^{19}$F. For interactions where tabulated cross section data were unavailable, the default \texttt{MCNP6} nuclear interaction models (\texttt{CEM03.03}) were used.

The \texttt{QGSP-BIC-allHPT} reference physics list was used in the \texttt{GEANT4} simulation. The cross section library used for proton interactions was \texttt{G4TENDL}, based on \texttt{TENDL-2014} \cite{rochman2012tendl}. For neutron cross sections and final states, \texttt{G4NDL}, based on \texttt{JEFF-3.3} ~\cite{plompen2020joint}, was used. All simulations were conducted with the ``\texttt{do\_not\_adjust\_final\_state}'' flag set to true.

The incident proton source was defined as a monoenergetic unidirectional beam with a diameter of 10 mm. Protons were sampled uniformly across the beam cross section and directed onto the centre of the LiF target. The proton energies of 1960, 2040, and 2310 keV were simulated to match the experimental conditions. No broadening of the beam in energy or shape was assumed. Protons were followed until their energy was below the threshold for the $^7$Li(p,n) reaction.

\begin{figure*}[!htb]
\centering
\begin{subfigure}[b]{0.32\textwidth}
\centering
\includegraphics[width=0.98\textwidth]{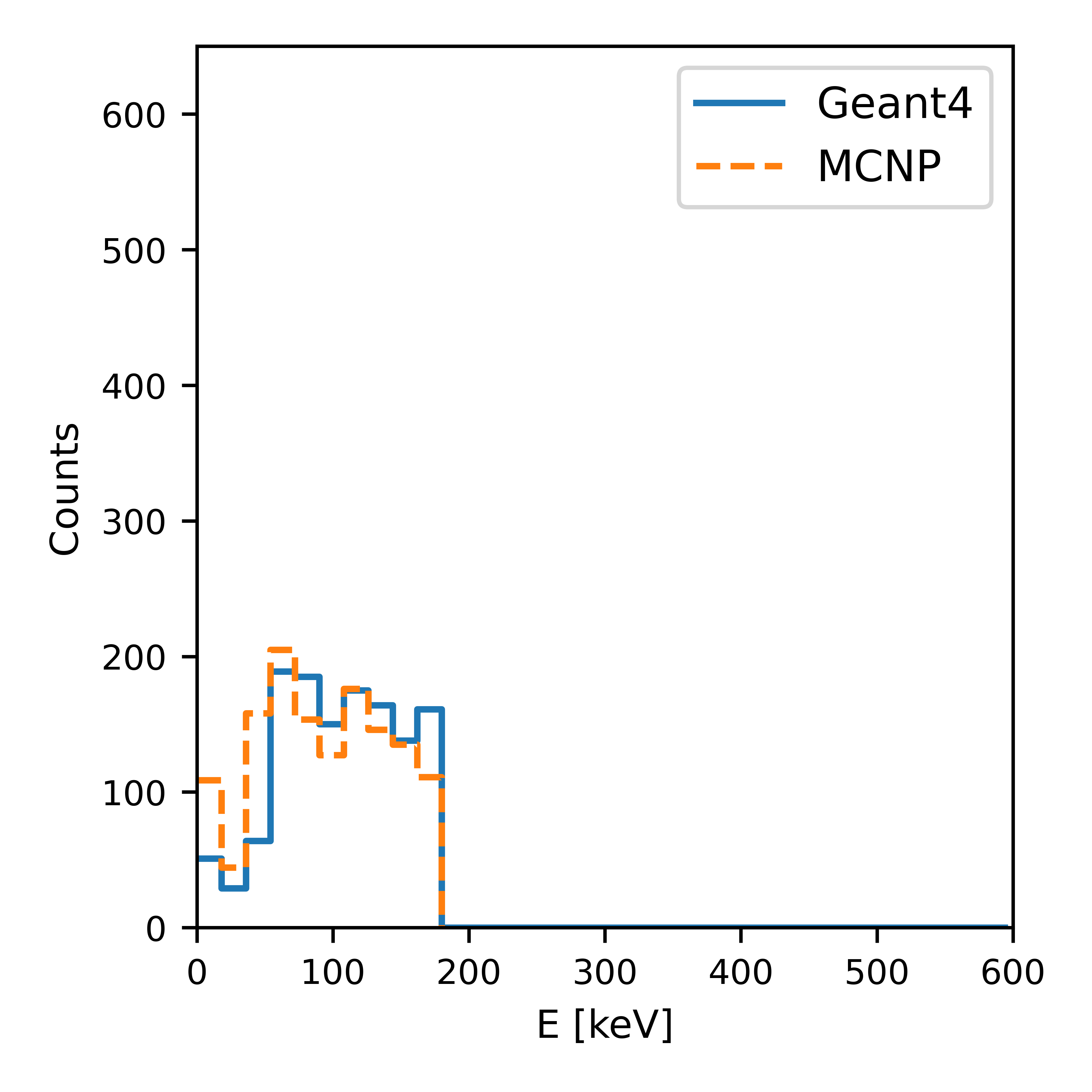}
\caption{}
\label{fig:MCNP_flux}
\end{subfigure}
\quad
\begin{subfigure}[b]{0.32\textwidth}
\centering
\includegraphics[width=0.98\textwidth]{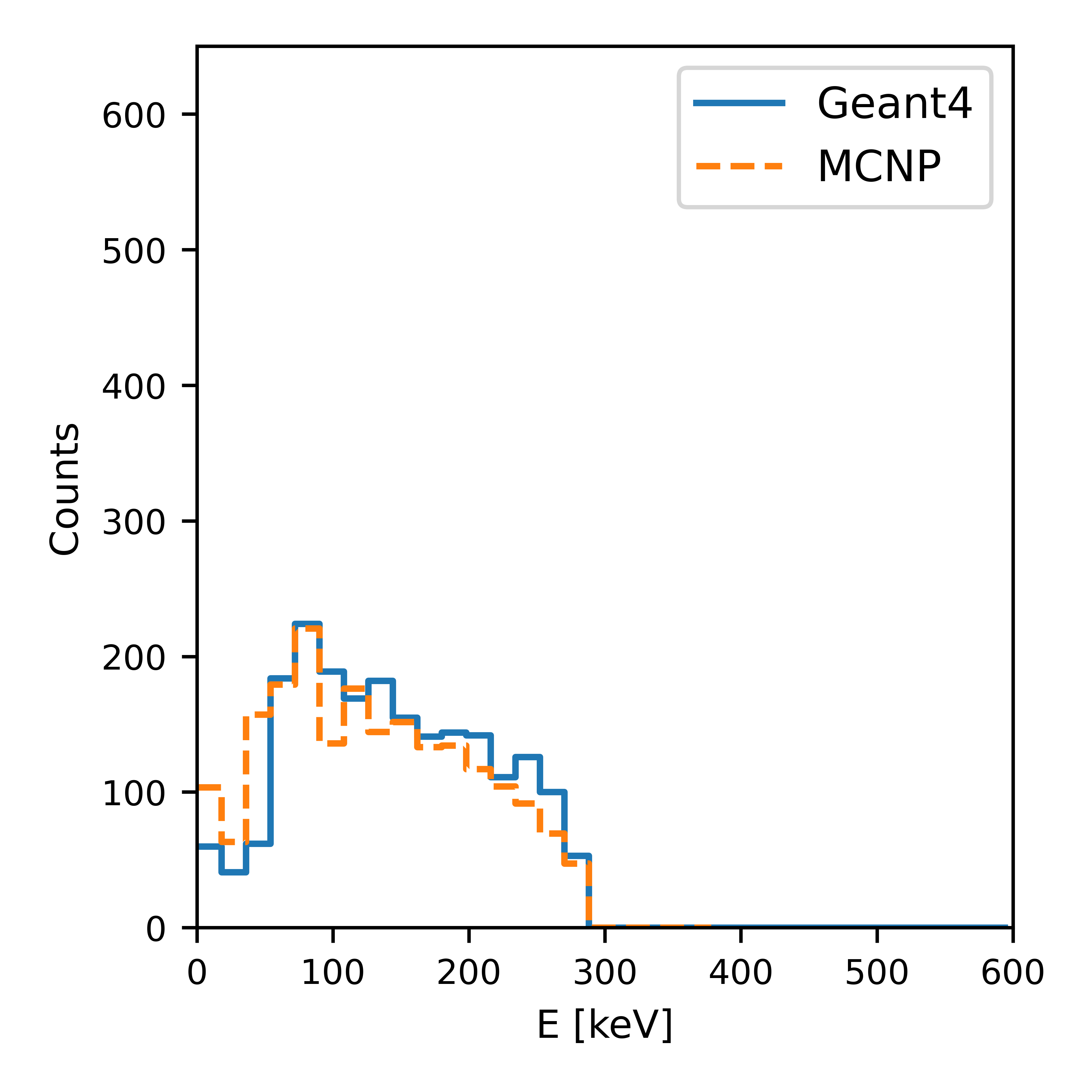}
\caption{}
\label{fig:2040_spectra}
\end{subfigure}
\begin{subfigure}[b]{0.32\textwidth}
\centering
\includegraphics[width=0.98\textwidth]{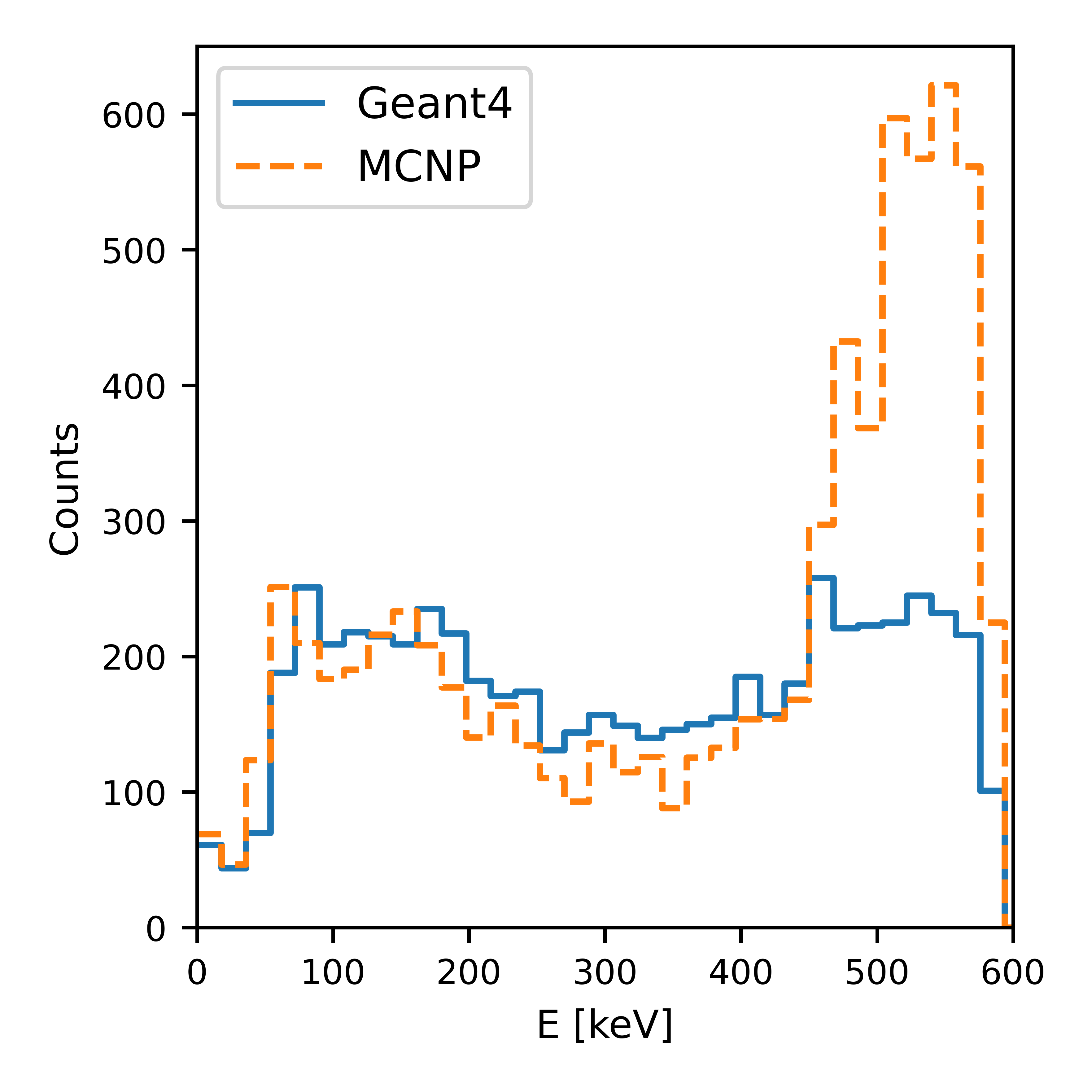}
\caption{}
\label{fig:2040_spectra}
\end{subfigure}
\caption{Simulated spectra with proton energy of 1960 keV (a), 2040 keV (b), and 2310 keV (c).}
\label{fig:simulated_spectra}
\end{figure*}

Neutrons were transported down to thermal energy and followed until they were tallied or escaped the simulation geometry. The neutron spectra were recorded by tallying neutrons crossing a circular plane with 10 cm diameter located 138 cm downstream of the centre of the target, corresponding to the collimating aperture of the NCT-WES spectrometer. The \texttt{F2} average surface flux tally was used in \texttt{MCNP6} to obtain the neutron energy spectra, while in \texttt{GEANT4} an equivalent approach was used. The simulated spectra are shown in \ref{fig:simulated_spectra}. The histograms have been re-binned to a fairly large bin of 18 keV to reduce statistical variance.

\section{NCT-WES}
\label{sec:nct-wes}
\subsection{Data elaboration before the unfolding}

The counts in the internal detectors of the spectrometer are given by Eq. \ref{eq_1}:
\begin{align}
    C_i&=B_\mathrm{target} F_{0^\circ} \int_0^\infty dE \cdot \varphi(E) \cdot R(138\,\mathrm{cm},\,E,\,i) \label{eq_1} \\
    &\mathrm{for} \quad i=1,\ldots,6 \nonumber
\end{align}
Where:

$\bm{C_i}$ is the count in the i-th NCT-WES internal detector (1= shallowest, 6 = deepest), normalised to the current on the target, after subtracting the shadow-cone measurements from the free-field one. Both shadow-cone and free-field measurements are normalised to the current on a target.

$\bm{B_\mathrm{target}}$ is the number of neutrons emitted by the target, normalised to the current on a target.

$\bm{F_{0^\circ}}$ is the anisotropy factor of the target in the forward direction ($0^\circ$) used for the measurements;

$\bm{\varphi(E)}$ is the energy distribution of the neutron field at the centre of the spectrometer front-face, normalised to unit fluence. This is also called the unit spectrum and gives the “quality” of the field (energy distribution), disregarding the “quantity” (fluence).

$\bm{R(138\,\mathrm{\textbf{cm}},\,E,\,i)}$ is the NCT-WES response matrix calculated for a source-to-spectrometer front-face distance of 138 cm, as applicable to this work, see Figure \ref{fig:sim_res_nct-wes}. 

\begin{figure}[h]
\centering
\includegraphics[width=0.48\textwidth]{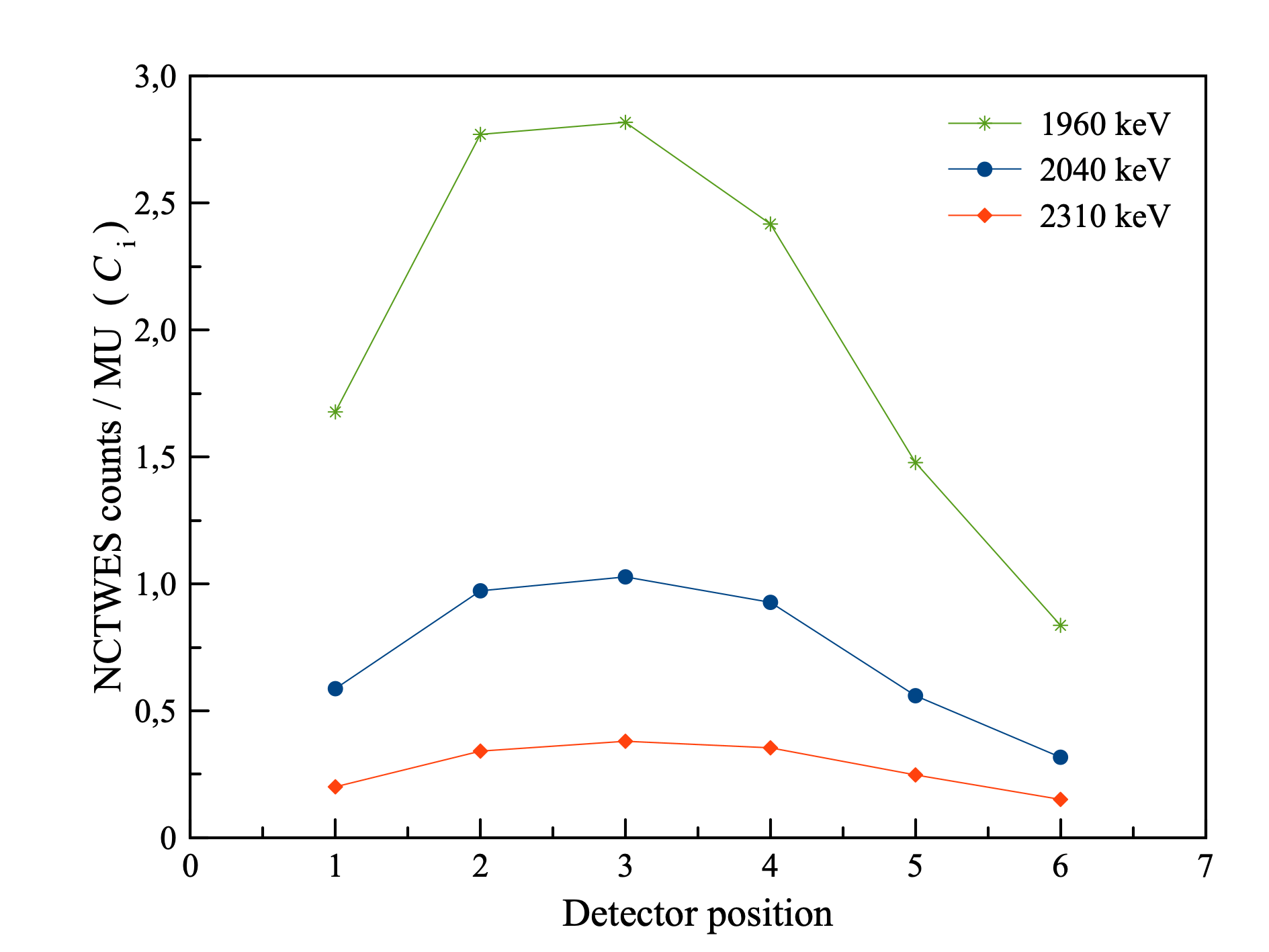}
\caption{$C_i$ represented as a function of the beam energy and detector position. Uncertainties (below 1\%-2\%) are smaller than the graphical symbols.}
\label{fig:C_i}
\end{figure}

$C_i$ is represented in Figure \ref{fig:C_i} as a function of beam energy and detector position. As explained elsewhere \cite{pietropaolo2020neutron}, these curves include all spectrometric information on neutron fields. The spectrum of target neutrons tends to harden as the proton energy increases. Similarly, the maximum in Figure \ref{fig:C_i} tends to shift to deeper positions. 
Spectrum unfolding in BS or SMNS is an under-determined problem, i.e., infinite mathematical functions exist which can reproduce the observed counts ($C_i$) when folded with a fixed response matrix. In addition, the resolving power of the spectrometer is limited by the shape of the response functions \cite{pietropaolo2020neutron}. Consequently, the user needs to ``feed'' the unfolding process with a given amount of pre-information, typically in the form of a reliable Monte Carlo spectrum. This ``guess spectrum'' is used to initiate the convergence process. In this work, two Monte Carlo spectra are available for each beam energy, calculated with \texttt{GEANT4} and \texttt{MCNP6} as described in Section \ref{sec:simulations}. A pragmatic approach to evaluate the plausibility/reliability of a guess spectrum, before the unfolding process, is that of comparing the experimental counts with those obtained by applying Eq.\,1 to that guess spectrum, after normalising every series to its average. Let us define:
\begin{align}
%    C_{i,\,\mathrm{exp}} = & \mathrm{\;NCT \text{-} WES\;counts\;normalised\;to}\\ &\mathrm{\;the\;charge\;on\;target},\nonumber\\
    P_{i,\,\mathrm{exp}} = & \;\frac{C_{i,\,\mathrm{exp}}}{<C_{i,\,\mathrm{exp}}>},\\
%    C_{i,\,\mathrm{guess}} = & \mathrm{\;Results\;from\;Eq.\,1\;applied\;to}\\&\mathrm{\;a\;given\;guess\;spectrum},\nonumber\\
    P_{i,\,\mathrm{guess}} = & \;\frac{C_{i,\,\mathrm{guess}}}{<C_{i,\,\mathrm{guess}}>},
\end{align}
where $C_{i,\,\mathrm{exp}}$ are the NCT-WES counts normalised to the current on target and $C_{i,\,\mathrm{guess}}$ the results of Eq.\,\ref{eq_1} applied to a given guess spectrum.
Under the idealised assumption that the guess spectrum is identical to the experimental one, $P_{i,\,\mathrm{exp}}$ and $P_{i,\,\mathrm{guess}}$ would coincide for every value of $i$. 

A quantitative way to assess the plausibility of a guess spectrum, before unfolding, is by studying the mean deviation $<d>=\overline{d}$ [24], defined as follows:
\begin{equation*}
    d_i=\frac{|P_{i,\,\mathrm{exp}}-P_{i,\,\mathrm{guess}}|}{u_i \cdot P_{i,\,\mathrm{exp}}},
\end{equation*}
where $i=1,\ldots ,6$ (i refers to the detector position) and $u_i$ is the relative uncertainty obtained by quadratically combining the uncertainty of the experimental data (1-2\%) and that of the response matrix (2\% \cite{bedogni2021nct}). When different guess spectra are derived on a physical basis, as in this work, it is common expert advice that the most plausible is the one which minimises $<d>$. According to this evaluation (see Table \ref{tab:deviation_to_MC}) the \texttt{MCNP6} simulation is more plausible than \texttt{GEANT4} for 1960 and 2040 keV, whilst \texttt{GEANT4} is more plausible than \texttt{MCNP6} for 2310 keV.
\begin{table}[t]
    \centering
    \caption{Average deviation < d > for the experimental data and the \texttt{MCNP6} or \texttt{GEANT4} guess spectra.}\label{tab:deviation_to_MC}
    \begin{tabular}{lccc}\toprule
        Proton energy [keV] & 1960 & 2040 & 2310 \\\midrule
        \texttt{MCNP6} & 1.2 & 1.3 & 2.1 \\
        \texttt{GEANT4} & 2.0 & 1.8 & 1.2 \\\bottomrule
    \end{tabular}
\end{table}

\subsection{Data unfolding}

The $C_{i, \mathrm{exp}}$ arrays obtained for the three proton energies were unfolded using the FRUIT code in special-gradient mode (FRUIT-SGM) \cite{bedogni2007fruit,amgarou2011measurement}. To better understand the roles of the pre-information and the resolving power of the spectrometer, three unfolding sessions were performed for every proton energy, using the \texttt{MCNP6} simulation, the \texttt{GEANT4} simulation, and a flat-in-lethargy (FIL) spectrum as guess spectra. Since the FIL spectrum does not convey any information about the neutron energy, the corresponding unfolding test enables an assessment of the spectrometer's intrinsic resolving power -- that is, its ability to distribute the neutron fluence as a function of energy in the absence of prior information. 

The unfolding results are presented in Figures \ref{fig:GEANT4_unfold_spectra}, \ref{fig:MCNP6_unfold_spectra} and \ref{fig:GEANT4_MCNP6_flat_unfold_spectra} and are explained below. Bin-by-bin uncertainties in the FRUIT results are derived from propagating the input uncertainties (NCT-WES counts and response matrix) through the unfolding process \cite{amgarou2011measurement}.

Figures \ref{fig:GEANT4_unfold_spectra:a} and \ref{fig:MCNP6_unfold_spectra:a} report the 1960 keV unfolding, starting from \texttt{GEANT4} or \texttt{MCNP6}, respectively. The guess spectra are also reported. The \texttt{GEANT4} simulation (Figure \ref{fig:GEANT4_unfold_spectra:a}) overestimates the higher energy region. As anticipated by the analysis of the average deviations $<d>$, the \texttt{MCNP6} guess is very similar to the result of the unfolding process, confirming the high plausibility of this guess. Figure \ref{fig:GEANT4_MCNP6_flat_unfold_spectra:a} compares the results of the unfolding process starting from \texttt{GEANT4}, \texttt{MCNP6}, and FIL. The three unfolded spectra show a very low average deviation (about 0.4), so they all reproduce the experimental counts well. The spectra obtained using \texttt{MCNP6} and \texttt{GEANT4} as initial guess spectra are nearly identical above 100 keV. Below this energy, the observed deviations reflect the limited resolving power of the technique. Nevertheless, the data in Table \ref{tab:deviation_to_MC} suggests considering the ``FRUIT from \texttt{MCNP6}''  as the best experimental estimation of the neutron spectrum from the 1960 keV proton beam. Interestingly, the ``FRUIT from flat-in-lethargy'' spectrum follows quite well the behaviour of other spectra, but clearly it cannot include fine structures because they are not present in the guess spectrum. 

Similar considerations apply to the 2040 keV proton beam (Figures \ref{fig:GEANT4_unfold_spectra:b}, \ref{fig:MCNP6_unfold_spectra:b} and \ref{fig:GEANT4_MCNP6_flat_unfold_spectra:b}), but in this case the unfolding results from \texttt{GEANT4} of \texttt{MCNP6} are considerably more coherent than in the 1960 keV case. Indeed, the ``FRUIT from \texttt{MCNP6}'' and ``FRUIT from \texttt{GEANT4}'' spectra (Figure \ref{fig:GEANT4_MCNP6_flat_unfold_spectra:b}) are nearly equivalent. 

For the 2310 keV case (Figures \ref{fig:GEANT4_unfold_spectra:c}, \ref{fig:MCNP6_unfold_spectra:c} and \ref{fig:GEANT4_MCNP6_flat_unfold_spectra:c}), the situation is reversed: as anticipated by the average‑deviation analysis, the \texttt{GEANT4} guess spectrum exhibits better agreement with the experimental data than the \texttt{MCNP6} guess spectrum. Consequently, the unfolded spectrum resembles more closely the \texttt{GEANT4} input than the \texttt{MCNP6} input. In this case, the ``FRUIT from \texttt{GEANT4}'' result can be considered to be the most plausible for the 2310 keV proton beam.

Table \ref{tab:max_energies} presents the maximum and mean energies of the unfolded spectra from both \texttt{GEANT4} and \texttt{MCNP6} guess spectra. For the ISO recommended monoenergetic reference fields from 1960, 2040 and 2310 keV protons, the mean energies and FWHMs are $144 \pm 12 $ keV, $250 \pm 11$ keV, and $565 \pm 8$ keV, respectively \cite{nolte2011monoenergetic}. As expected, the proton energy loss across the thick target volume broadens the neutron spectrum, resulting in lower overall mean energy.

\begin{table*}[t]
    \centering
    \caption{Maximum and mean neutron energies of the measured spectra, based on the unfolded data using \texttt{GEANT4} and \texttt{MCNP6} guess spectra.}\label{tab:max_energies}
    \begin{tabular}{ccccc}\toprule
        Beam energy & $E_{\text{max}}$, \texttt{GEANT4} guess & $E_{\text{max}}$, \texttt{MCNP6} guess & $E_{\text{mean}}$, \texttt{GEANT4} guess & $E_{\text{mean}}$, \texttt{MCNP6} guess \\\midrule
        1960\,keV& 181 & 179 & 90 & 89 \\
        2040\,keV & 281 & 281 & 120 & 119 \\
        2310\,keV & 587 & 587 & 278 & 295 \\\bottomrule
    \end{tabular}
\end{table*}

\subsection{Determination of the fluence}

By inverting Eq.\,\ref{eq_1}, the unfolding process determines $\varphi(E)$ (Section 5) and the product $B_\mathrm{target}\cdot F_{0^\circ}$. The fluence per current on a target at the reference position, i.e.\, the centre of the NCT-WES front face, is,
\begin{equation*}
    \Phi_\mathrm{ref}=\frac{B_\mathrm{target}\cdot F_{0^\circ}}{4\,\pi\,r^2},
\end{equation*}
where $r = 138$\,cm. 

Table \ref{tab:determination_of_phi_ref} reports $\Phi_\mathrm{ref}$ for the studied neutron fields, based on the unfolded data using \texttt{GEANT4} or \texttt{MCNP6} guess spectra.
\begin{table*}[t]
    \centering
    \caption{Determination of $\Phi_\mathrm{ref}$ for the studied neutron fields, based on the unfolded data using \texttt{GEANT4} or \texttt{MCNP6} guess spectra. Uncertainties are in the order of 3\%, due to the quadratic combination of the statistical uncertainty from the unfolding process with that on the NCT-WES calibration factor (2.5\%).}\label{tab:determination_of_phi_ref}
    \begin{tabular}{ccc}\toprule
        & \multicolumn{2}{c}{$\Phi_\mathrm{ref}$\,(cm$^{-2}\mu \mathrm{A}^{-1}$)}\\
        Beam energy & \texttt{GEANT4} guess & \texttt{MCNP6} guess\\\midrule
        1960\,keV& $224\pm 7$ & $225\pm 7$ \\
        2040\,keV & $522 \pm 16$ & $523 \pm 16$  \\
        2310\,keV & $1520 \pm 46$ & $1515\pm 45$ \\\bottomrule
    \end{tabular}
\end{table*}

\section{Berthold LB 6411}
\label{sec:berthold}

The cps value was periodically read from the camera feed. The mean of all recorded values was used as the final cps value for each measurement.

\begin{table*}[t]
    \centering
    \caption{Data of the irradiation for the LB 6411 measurements: mean proton current, measurement times, and variability range for the scattered fraction $s$.}\label{tab:exp_data_summary_berthold}
    \begin{tabular}{lccc}\toprule
        Proton energy [keV] & 1960 & 2040 & 2310 \\\midrule
        Proton current, free field [$\mu$A]& 0.236 $\pm$ 0.009 & 0.229 $\pm$ 0.008 & 0.293 $\pm$ 0.008\\
        Proton current, shadow cone [$\mu$A]& 0.145 $\pm$ 0.006 & 0.301 $\pm$ 0.005 & 0.365 $\pm$ 0.007\\
        Measurement time, free field (s) & 300 & 300 & 300 \\
        Measurement time, shadow cone (s) & 300 & 300 & 300 \\
        Scattered fraction $s$ & 2.7\% & 8.0\% & 5.6\%
        \\\bottomrule
    \end{tabular}
\end{table*}

The LB 6411 rem counter has been designed to have a response function $R(E)$, such that the detector response is proportional to the ambient dose equivalent $H_{10}^*$
\begin{equation*}
    C = \Phi \cdot \int{R(E) \cdot \varphi(E) \cdot dE},
\end{equation*}
where $\Phi$ is the total flux at the surface of the detector, and $\varphi(E)$ the unit spectrum. The total flux can then be estimated by
\begin{equation*}
      \Phi = \frac{C}{\int{R(E) \cdot \varphi(E) \cdot dE}},
\end{equation*}
for a given spectrum. 

\begin{figure}[!htb]
\centering
\includegraphics[width=0.43\textwidth]{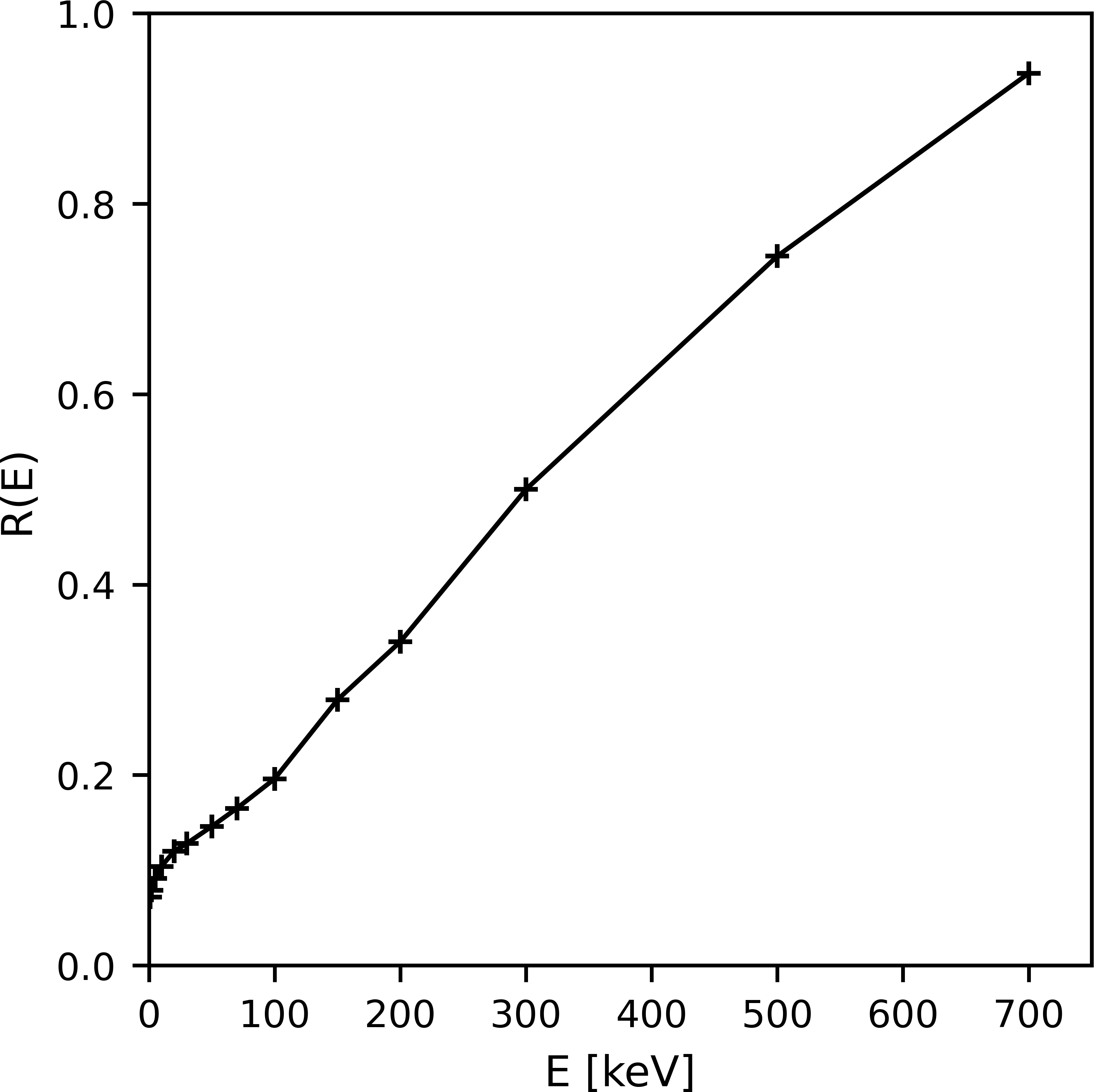}
\caption{Response function of LB 6411 from Reference~\cite{Burgkhardt1997}.}
\label{fig:response}
\end{figure}

The flux was calculated by using the response function given in \cite{Burgkhardt1997} and the simulated spectra. The response function is presented in Figure~\ref{fig:response}. The response of each energy bin in the unit spectrum was calculated independently. The response was estimated by linear interpolation between the data points. The main uncertainties arose from the beam current measurements (3--4\,\%) and the response conversion factor (1--3\,\%).

Table \ref{tab:bertholdi_phi_ref} presents the neutron fluences calculated from the measurements with the LB 6411 monitor.

It has been observed that the count rate measured with the LB 6411 has a tendency to gradually increase for up to 8\,\% during the first 15 $\mu$Sv of irradiation. The source of this behaviour is not known, but is thought to arise due to charging-up effects inside the \ce{^{3}He} tube. This so-called warm-up effect was estimated to cause a further 5\,\% uncertainty to the cps measurement. The estimated uncertainties of the shadow cone corrected fluxes, including the warm-up effect, were 6--8\,\%. 

\begin{table*}[t]
    \centering
    \caption{Fluences determined by Berthold using proportionality factors calculated with  \texttt{GEANT4} and  \texttt{MCNP6} spectra. Uncertainties are in the order of 7~\%. }\label{tab:bertholdi_phi_ref}
    \begin{tabular}{ccc}\toprule
        & \multicolumn{2}{c}{$\Phi$\,(cm$^{-2}\mu \mathrm{A}^{-1}$)}\\
        Beam energy & \texttt{GEANT4} & \texttt{MCNP6} \\\midrule
        1960\,keV& $130\pm 10$ & $138\pm 11$ \\
        2040\,keV & $350 \pm 30$ & $380 \pm 30$  \\
        2310\,keV & $1430 \pm 90$ & $1250\pm 70$ \\\bottomrule
    \end{tabular}
\end{table*}

\section{Discussion \& Conclusions}
\label{sec:conclusions}

\begin{figure*}[!htb]
\centering
\begin{subfigure}[b]{0.42\textwidth}
\centering
\includegraphics[width=0.98\textwidth]{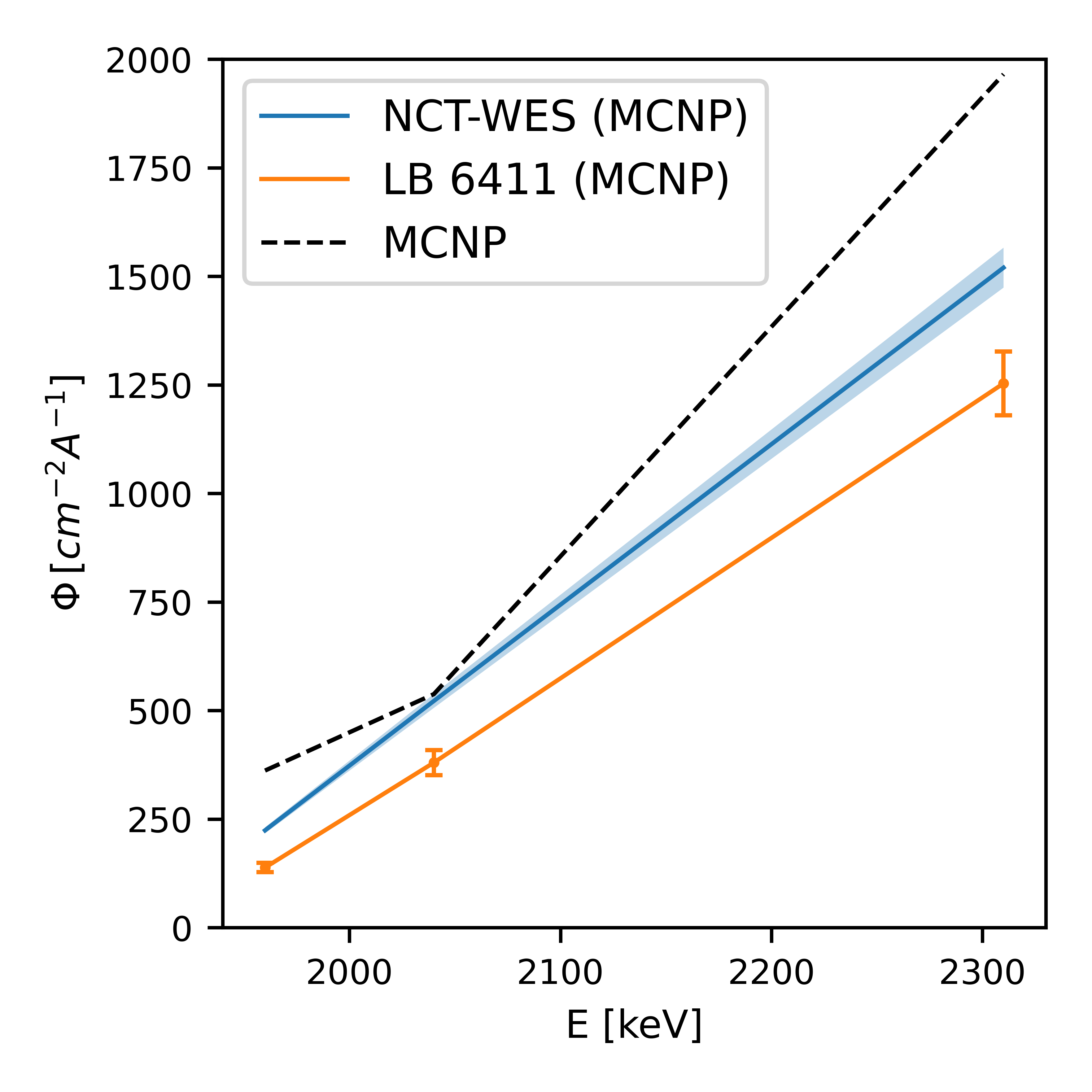}
\caption{}
\label{fig:MCNP_flux}
\end{subfigure}
\quad
\begin{subfigure}[b]{0.42\textwidth}
\centering
\includegraphics[width=0.98\textwidth]{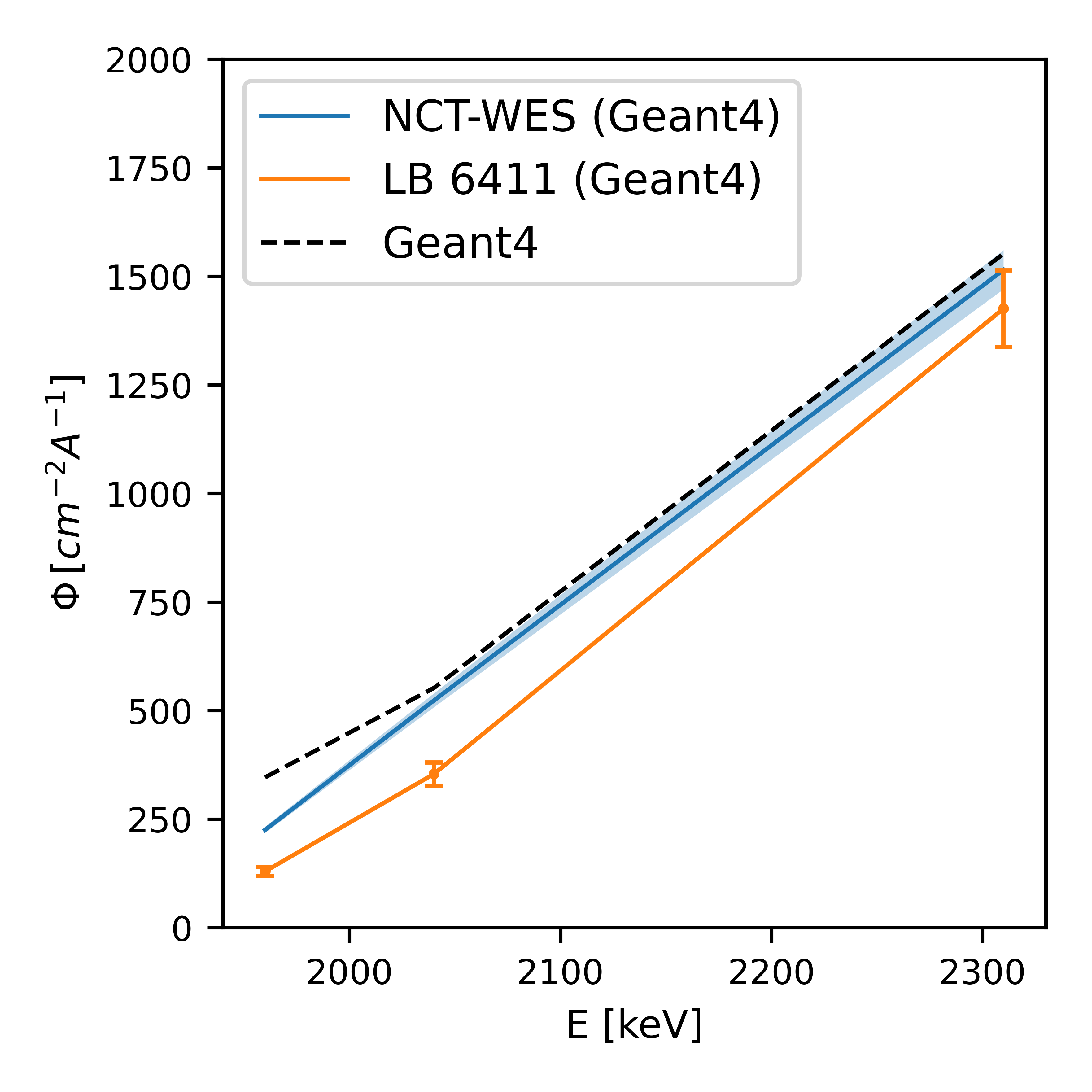}
\caption{}
\label{fig:Geant4_flux}
\end{subfigure}
\caption{$\Phi$ values determined using \texttt{MCNP6} (a), and \texttt{Geant4} (b) as the ground truth.}
\end{figure*}

The neutron beams in the forward direction from the University of Helsinki LiF (p,n) production target were characterized using the NCT-WES single-moderator spectrometer and independent flux measurements. The room scattering was subtracted using the shadow-cone technique. The plausibility of the guess spectra simulated with \texttt{GEANT4} and \texttt{MCNP6} was assessed through the analysis of average deviations $<d>$. \texttt{MCNP6} was found to be more plausible for the 1960 and 2040 keV beams, and \texttt{GEANT4} for the 2310 keV beam. The unfolding process confirmed these hypotheses, producing unfolded spectra that are very similar to the \texttt{MCNP6} guess (for 1960 and 2040 keV) and to the \texttt{GEANT4} guess (for 2310 keV). 

The fluence per current on a target at the spectrometer front face was derived for every beam with uncertainty of the order of $\pm 3\,\%$, finding very little or no dependence from the guess spectra used for the unfolding. The whole measurement campaign took about 15 hours of beam time, and the scatter fraction was found to be lower than about 3\,\%. NCT-WES proved to be a very convenient instrument for the characterisation of directional neutron irradiation facilities.

A comparison of Tables \ref{tab:determination_of_phi_ref} and \ref{tab:bertholdi_phi_ref} shows that the choice of a simulation code leads to a larger relative discrepancy in the neutron fluence obtained using the LB 6411 monitor for all proton energies, while the fluences measured with the NCT-WES show greater consistency between \texttt{MCNP6} and \texttt{GEANT4}. Since the fluence obtained with the LB 6411 is calculated using the response function of the device and the simulated spectra, the shape of the spectrum directly affects the result. Thus, it is important that the simulated spectra accurately describe the neutron field, if the LB 6411 is used for beam monitoring.

Figure \ref{fig:MCNP_flux} presents the neutron fluences from the \texttt{MCNP6} simulations, and the fluences determined by the NCT-WES spectrometer and the Berthold LB 6411 monitor. The \texttt{MCNP6} spectra were used as guess spectra for the NCT-WES unfolding and to calculate the LB 6411 fluence as explained in Sections \ref{sec:nct-wes} and \ref{sec:berthold}. The corresponding results for \texttt{GEANT4} are shown in Figure \ref{fig:Geant4_flux}. The fluence measured with the LB 6411 monitor is systematically lower than the fluence determined by NCT-WES for each proton energy, however, this difference can be corrected with a suitable calibration factor. 

Both simulation tools overestimate the neutron fluence for the lowest proton energy, while \texttt{MCNP6} models the spectrum shape more accurately. For the 2040 keV proton beam, both \texttt{MCNP6} and \texttt{GEANT4} model the fluence and shape of the spectrum rather accurately. The largest differences between \texttt{MCNP6} and \texttt{GEANT4} are seen in the results for the 2310 keV proton beam, which \texttt{GEANT4} models more accurately based on both the NCT-WES and LB 6411 results. \texttt{MCNP6} greatly overestimates the neutron flux in the energy range between 450 and 600 keV (Figure \ref{fig:simulated_spectra}). Similar differences in spectrum shape and neutron yield have previously been reported in \cite{macleod2026}. The differences between the two Monte Carlo codes likely result from different cross section libraries and physics models used in the simulations. Future work is planned to explore the discrepancies between the two transport codes. Both \texttt{MCNP6} and \texttt{GEANT4} model the neutron production from 1960 keV and 2040 keV protons reasonably well, whereas \texttt{GEANT4} should be used to model the neutron production from 2310 keV protons.

\section*{Data Availability Statement}
Data will be made available by the corresponding author on a reasonable request.

\section*{Acknowledgments}
The research presented in this paper was supported by the University of Helsinki, the Helsinki Institute of Physics and by INFN project nKEEP (Commissione Scientifica Nazionale CSN5 of INFN, Italy).

\bibliographystyle{elsarticle-num}
\bibliography{references}
\clearpage
\onecolumn
\appendix
\section{Unfolding results}

\begin{figure*}[!htb]
\centering
\begin{subfigure}[b]{0.32\textwidth}
\centering
\includegraphics[width=0.98\textwidth]{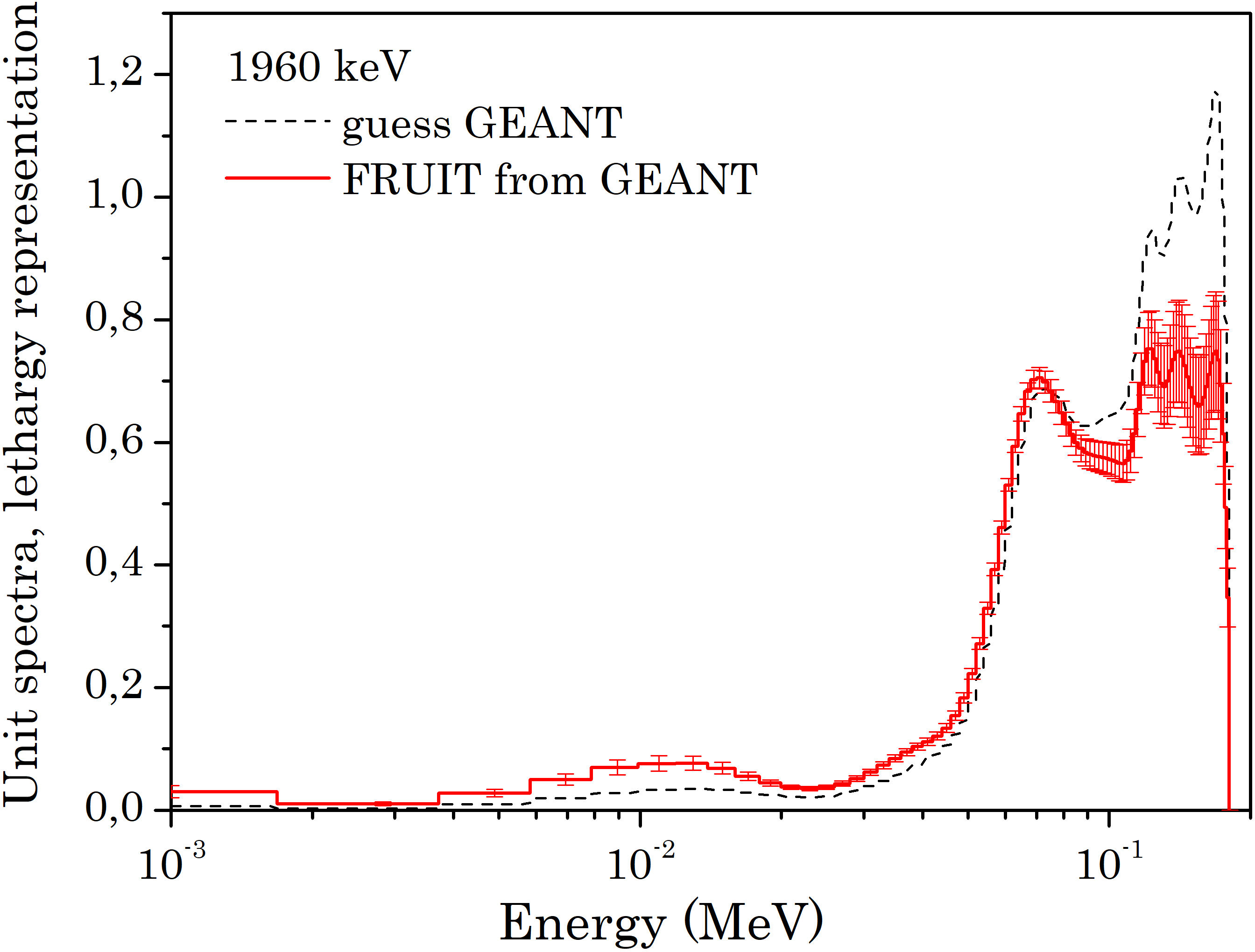}
\caption{1960 keV beam}\label{fig:GEANT4_unfold_spectra:a}
\end{subfigure}
\begin{subfigure}[b]{0.32\textwidth}
\centering
\includegraphics[width=0.98\textwidth]{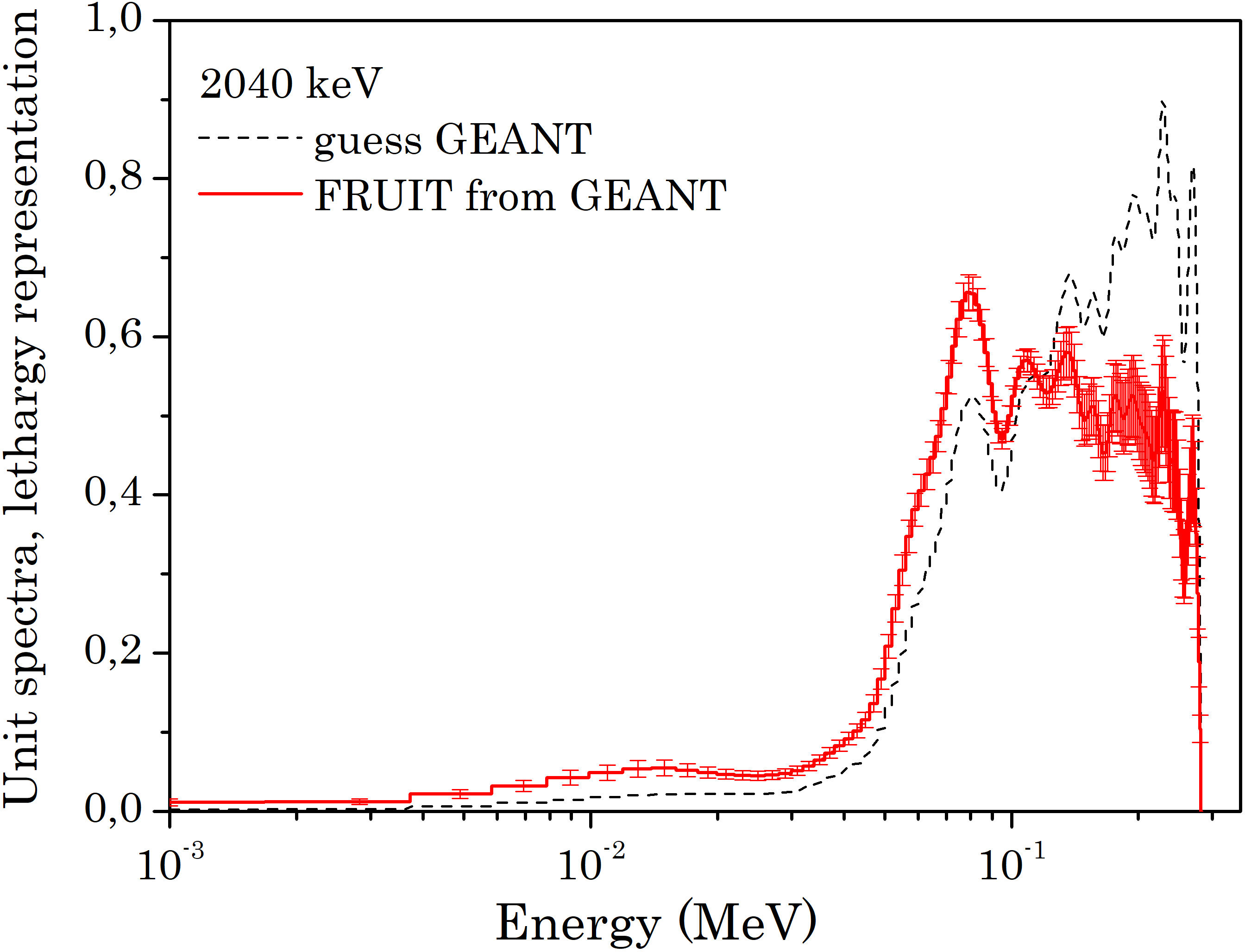}
\caption{2040 keV beam}\label{fig:GEANT4_unfold_spectra:b}
\end{subfigure}
\begin{subfigure}[b]{0.32\textwidth}
\centering
\includegraphics[width=0.98\textwidth]{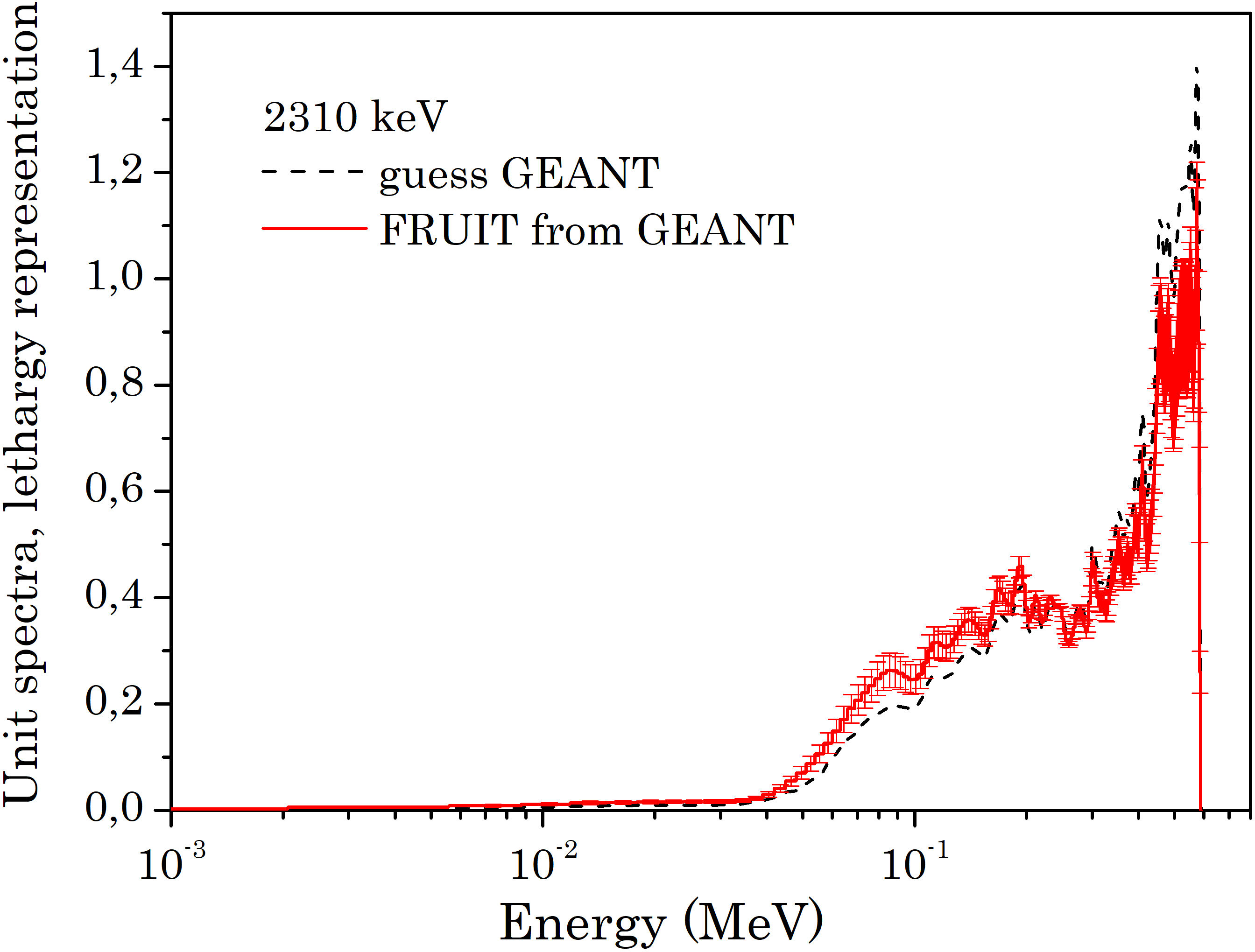}
\caption{2310 keV beam}\label{fig:GEANT4_unfold_spectra:c}
\end{subfigure}
\caption{\texttt{GEANT4} simulation and unfolded spectra, using \texttt{GEANT4} as a guess. The spectra are normalised to the unit fluence and are in equi-lethargy representation.}
\label{fig:GEANT4_unfold_spectra}
\end{figure*}
\begin{figure*}[!htb]
\centering
\begin{subfigure}[b]{0.32\textwidth}
\centering
\includegraphics[width=0.98\textwidth]{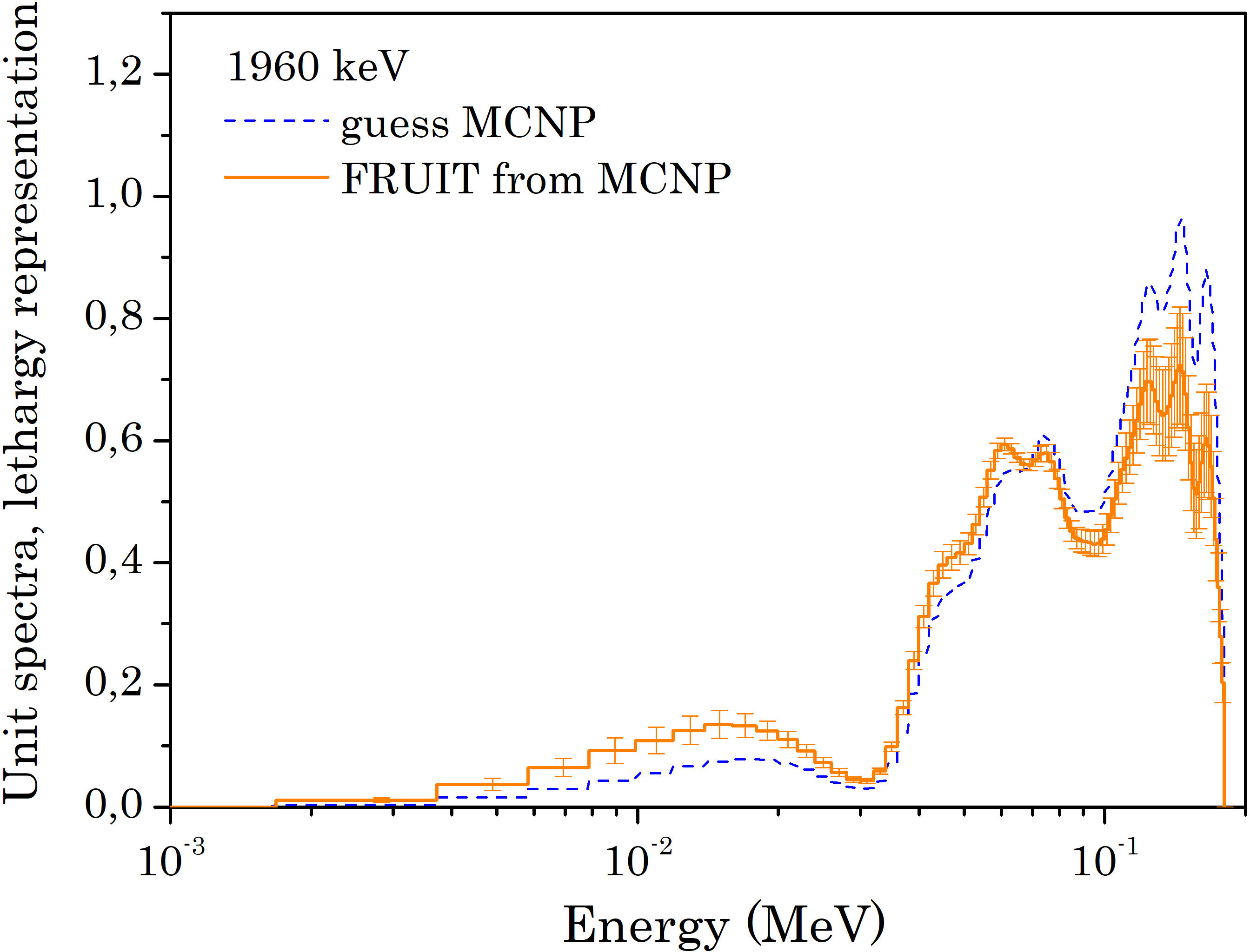}
\caption{1960 keV beam}\label{fig:MCNP6_unfold_spectra:a}
\end{subfigure}
\begin{subfigure}[b]{0.32\textwidth}
\centering
\includegraphics[width=0.98\textwidth]{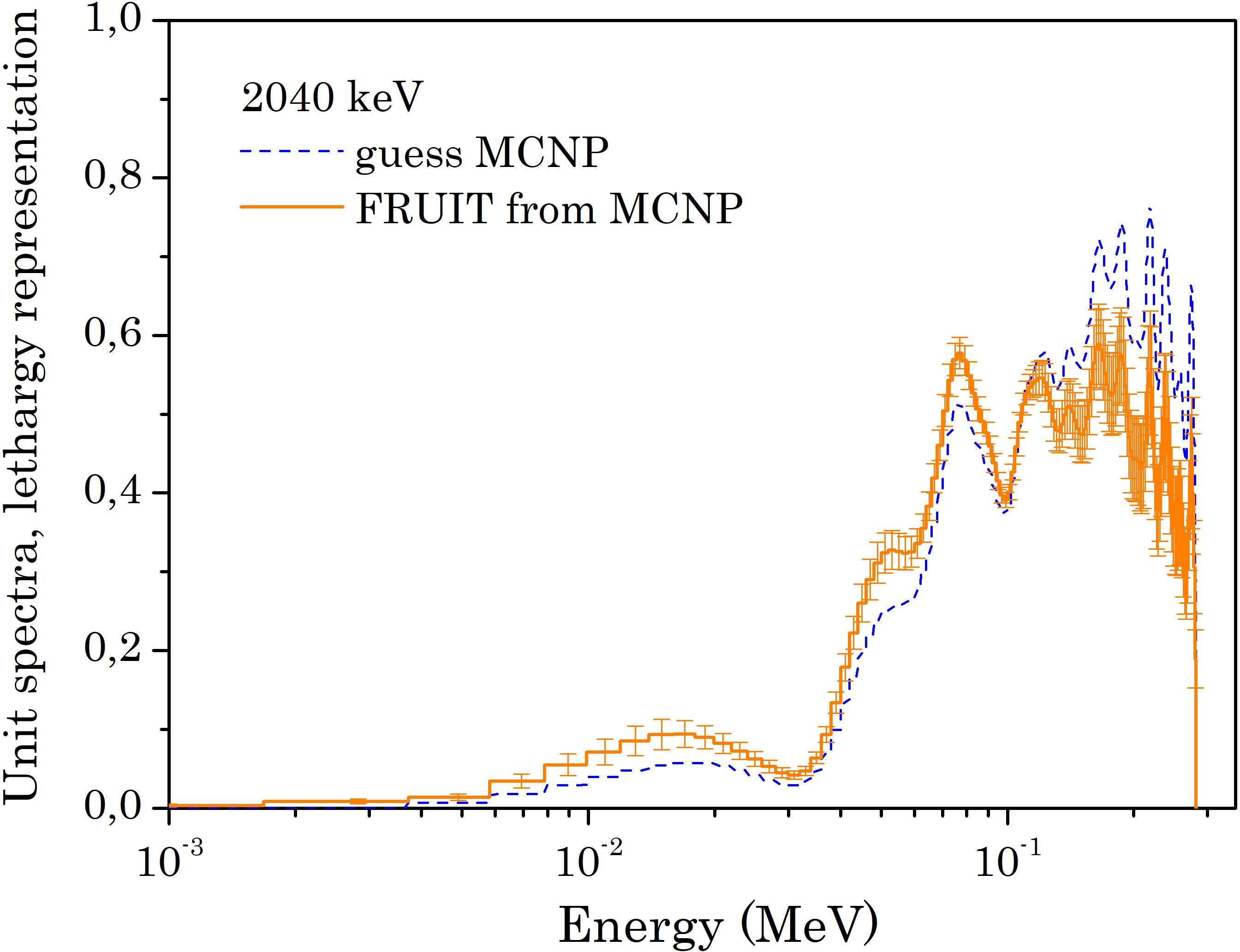}
\caption{2040 keV beam}\label{fig:MCNP6_unfold_spectra:b}
\end{subfigure}
\begin{subfigure}[b]{0.32\textwidth}
\centering
\includegraphics[width=0.98\textwidth]{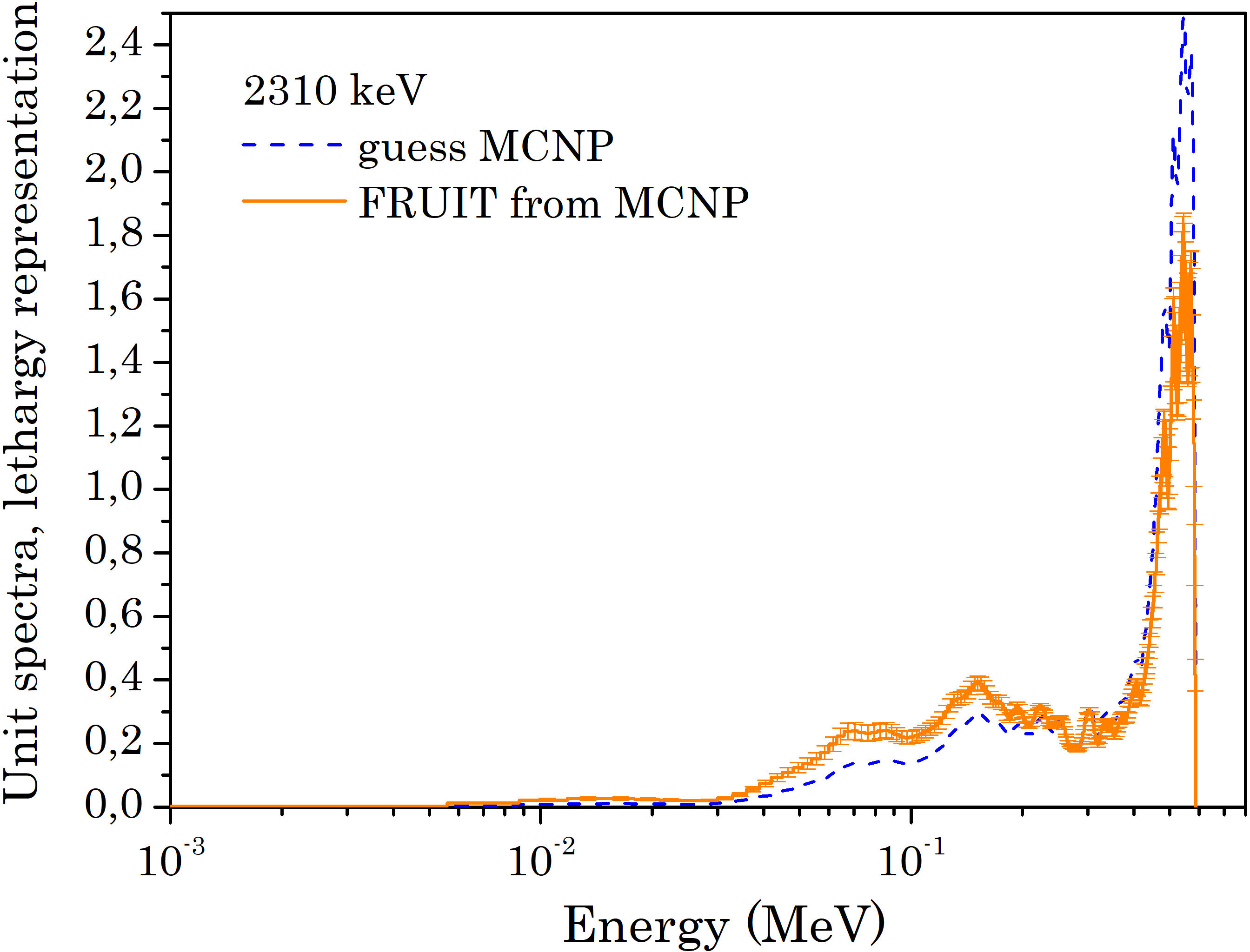}
\caption{2310 keV beam}\label{fig:MCNP6_unfold_spectra:c}
\end{subfigure}
\caption{\texttt{MCNP6} simulation and unfolded spectrum, using \texttt{MCNP6} as guess. The spectra are normalised to the unit fluence and are in equi-lethargy representation.}
\label{fig:MCNP6_unfold_spectra}
\end{figure*}
\begin{figure*}[!htb]
\centering
\begin{subfigure}[b]{0.32\textwidth}
\centering
\includegraphics[width=0.98\textwidth]{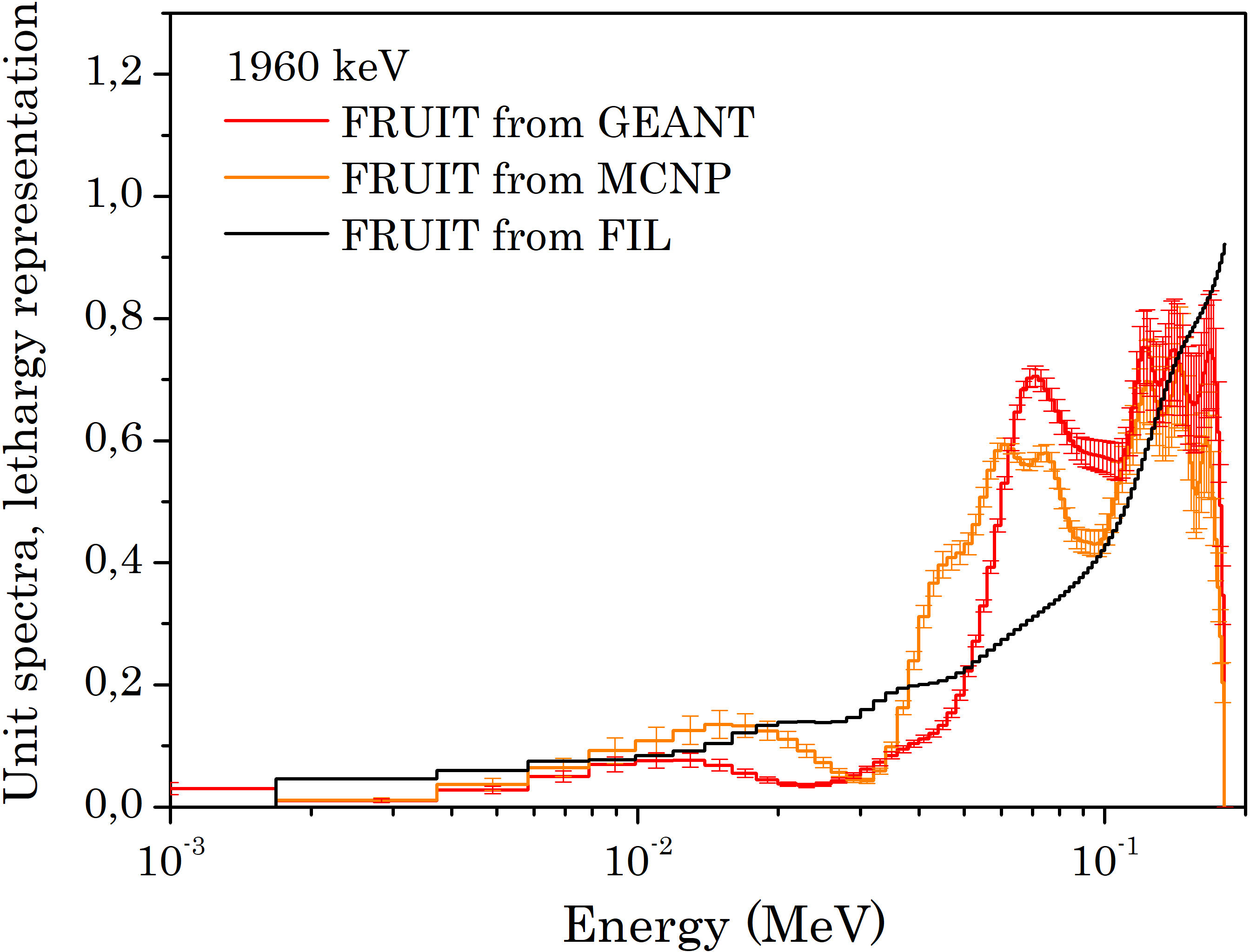}
\caption{1960 keV beam}\label{fig:GEANT4_MCNP6_flat_unfold_spectra:a}
\end{subfigure}
\begin{subfigure}[b]{0.32\textwidth}
\centering
\includegraphics[width=0.98\textwidth]{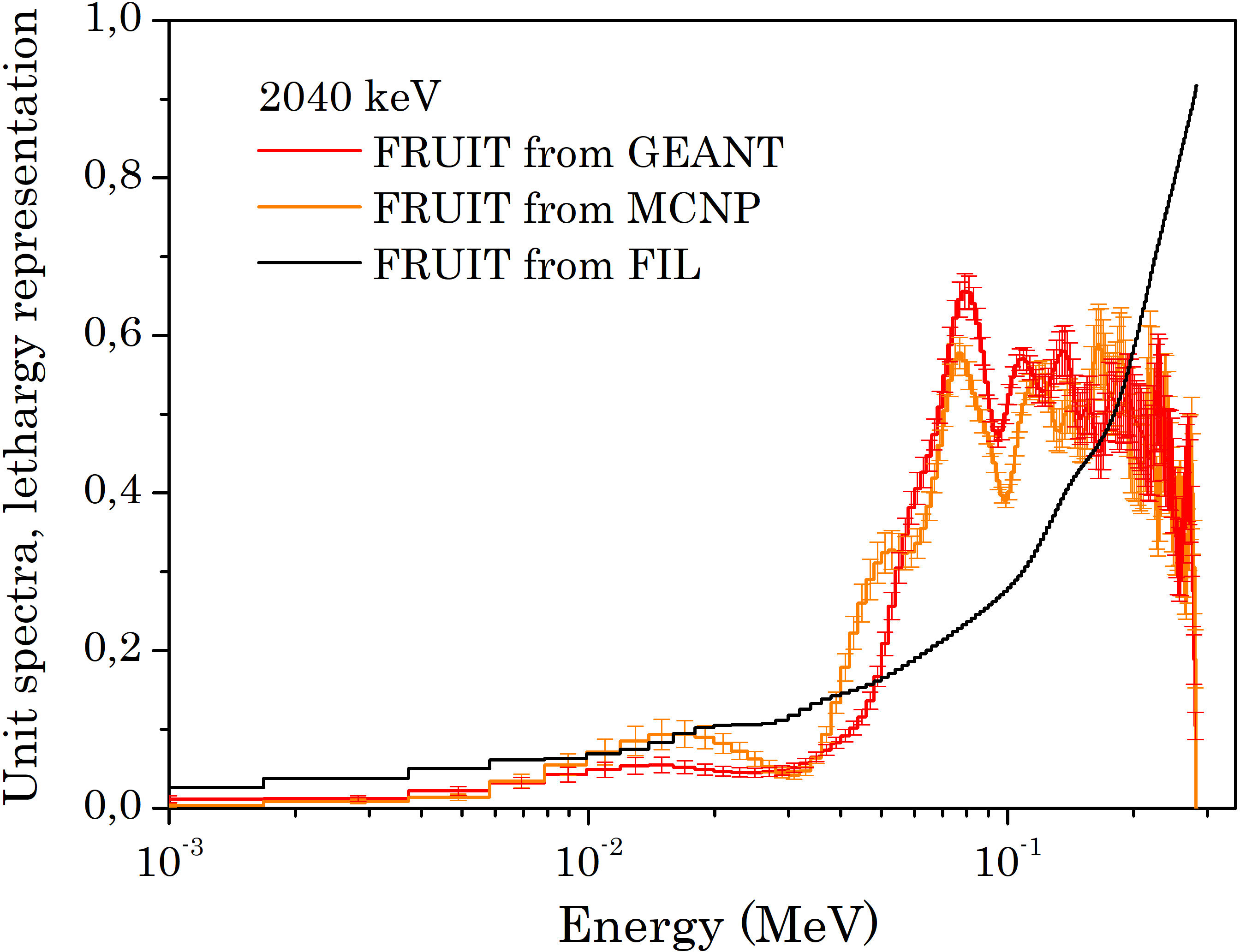}
\caption{2040 keV beam}\label{fig:GEANT4_MCNP6_flat_unfold_spectra:b}
\end{subfigure}
\begin{subfigure}[b]{0.32\textwidth}
\centering
\includegraphics[width=0.98\textwidth]{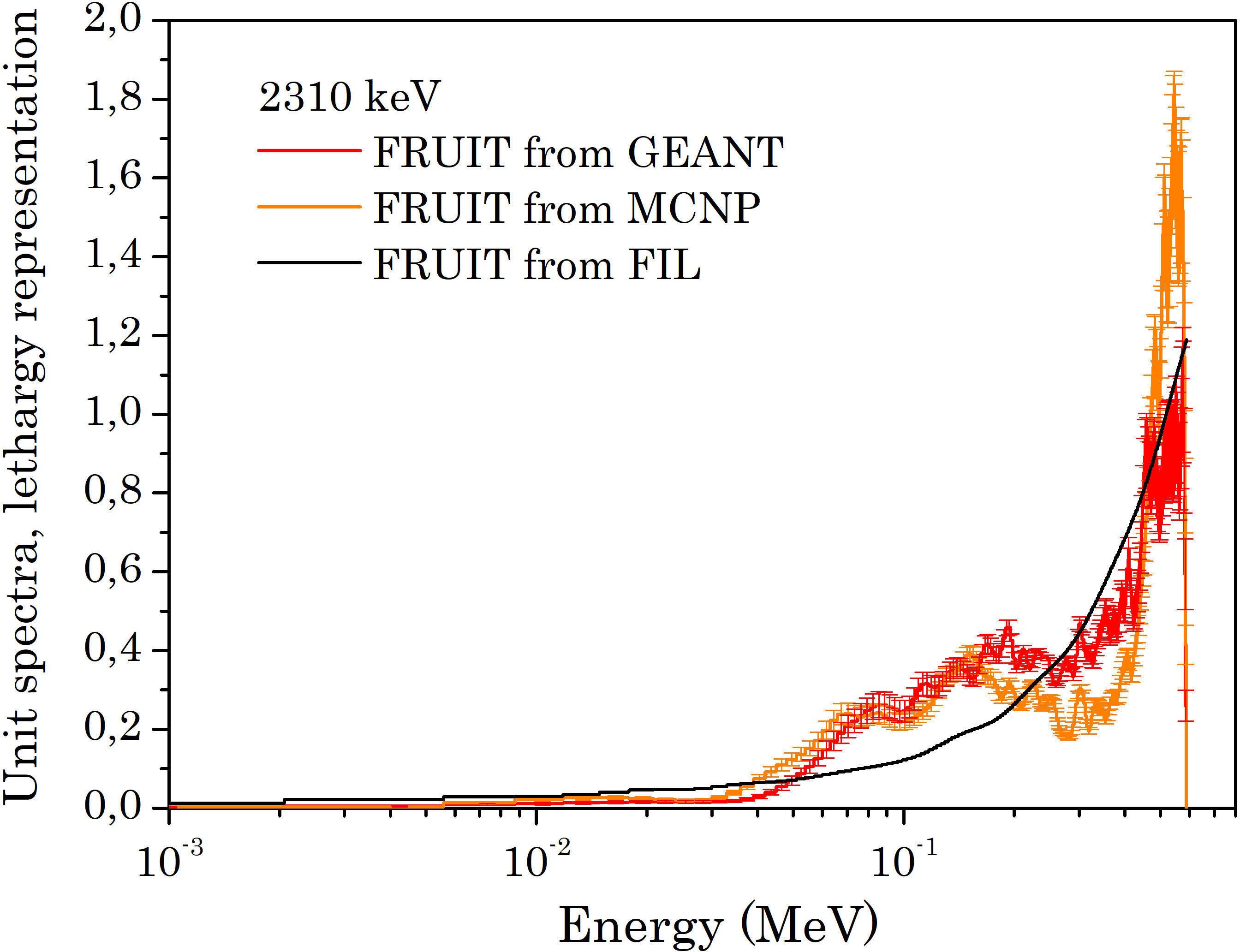}
\caption{2310 keV beam}\label{fig:GEANT4_MCNP6_flat_unfold_spectra:c}
\end{subfigure}
\caption{Unfolded spectra, starting from \texttt{GEANT4}, \texttt{MCNP6} and flat-in-lethargy. The spectra are normalised to the unit fluence and are in equi-lethargy representation.}
\label{fig:GEANT4_MCNP6_flat_unfold_spectra}
\end{figure*}

\end{document}